\documentclass[prd,a4paper,nofootinbib,
notitlepage,
]{revtex4-1}
\usepackage{graphicx}
\usepackage{bbold}
\usepackage{mathtools}
\usepackage{amsthm}
\usepackage{amsmath}
\usepackage{amsfonts}
\usepackage{amssymb}
\usepackage{slashed} 
\usepackage{color}
\usepackage{xcolor}
\usepackage{tabularx}
\usepackage{hyperref}
\usepackage{bm}% bold math
\usepackage{ulem}
\def\eq#1{{Eq.\,(\ref{#1})}}

\def\Re{\mbox{Re}\,}

\def\di{\mbox{d}}

\colorlet{grayline}{gray!70}
\definecolor{blueline}{rgb}{0,0.27,0.55}
\definecolor{Gray}{gray}{0.9}
\definecolor{oucrimsonred}{rgb}{0.6, 0.0, 0.0}
\definecolor{persianblue}{rgb}{0.11, 0.22, 0.73}
\definecolor{forestgreen}{rgb}{0.13,0.35,0.13}
 \hypersetup{colorlinks, citecolor=forestgreen, linkcolor=persianblue, urlcolor=forestgreen}
\def\hhref#1{\href{http://arxiv.org/abs/#1}{#1}} % in bibliography
\newcommand{\be}{\begin{equation}}
\newcommand{\ee}{\end{equation}}
\newcommand{\bea}{\begin{eqnarray}}
\newcommand{\eea}{\end{eqnarray}}
\newcommand{\nn}{\nonumber}

\newcommand{\com}[1]{}
\newcommand{\gsim}{\lower.7ex\hbox{$\;\stackrel{\textstyle>}{\sim}\;$}}
\newcommand{\lsim}{\lower.7ex\hbox{$\;\stackrel{\textstyle<}{\sim}\;$}}
\newcommand{\BR}{\mbox{BR\,}}
\newcommand{\bc}{\begin{center}}
\newcommand{\ec}{\end{center}}

\newcommand{\Rechi}{\Re\!\!\left[\chi(m^2_{\tau\bar{\tau}})\right]}
\newcommand{\Abschi}{\left|\chi(m^2_{\tau\bar{\tau}})\right|}
\newcommand{\mtautau}{m_{\tau\bar{\tau}}}
\begin{document}
%%%%%%%%%%%%%%%%%%%%%%%%%%%%%%%%%%%%%%%%%%%%%%%%%%%%%%%%%%%  FRONT PAGE

\title{Constraining new physics in entangled two-qubit systems:
  top-quark, tau-lepton and photon pairs}

\date{\today} 
\author{{\color{persianblue} M.~Fabbrichesi}$^{\dag}$}
\author{{\color{persianblue} R.~Floreanini}$^{\dag}$}
\author{{\color{persianblue} E.~Gabrielli}$^{\ddag\dag *}$}
\affiliation{\color{darkgray}$^{\dag}$INFN, Sezione di Trieste, Via  Valerio 2, 34127 Trieste, Italy }
\affiliation{$^{\ddag}$Physics Department, University of Trieste, Strada Costiera 11, 34151 Trieste, Italy}
\affiliation{$^{*}$Laboratory of High-Energy and Computational Physics, NICPB, Ravala 10, 10143 Tallinn, Estonia}

\begin{abstract}
\noindent The measurement of quantum entanglement can  provide a new and  most sensitive probe to physics beyond the Standard Model.  
We  use the concurrence of the top-quark pair spin states produced at colliders
to constrain the magnetic dipole term in the coupling between top quark and gluons, that of $\tau$-lepton pairs spin states to bound contact interactions and  that of $\tau$-lepton pairs or two-photons spin states from the decay of the Higgs boson
in trying to distinguish between  CP-even and odd couplings. 
These four examples show the power of the new approach as well as its limitations.
We show that differences in the entanglement  in the top-quark and 
$\tau$-lepton pair production cross sections  can provide constraints  better than those previously estimated from  total cross sections or classical correlations. Instead, the final states in the decays of the Higgs boson remain maximally entangled even in the presence of CP-odd couplings and cannot be used to set bounds on new physics.  We  discuss the violation of Bell inequalities featured in all  four processes. 
 \end{abstract} 

%\keywords{}
%\pacs{}
%%%%%%%%%%%%%%%%%%%%%%%%%%%%%%%%%%%%%%%%%%%%%%%%%%%%%%%%%%%%%%%%%%%
\maketitle 
%%%%%%%%%%%%%%%%%%%%%%%%%%%%%%%%%%%%%%%%%%%%%%%%%%%%%%%%%%%%%%%%%%%
\section{Entanglement at work in high-energy physics}

Collider physics is all about the production of particles from the interaction and the decay of other particles. Consider the case of the production of just two of these particles: We  can study  their momenta, energies and spin to reconstruct their properties and compare them to those expected in the Standard Model (SM)---or in an extension of it.  The same observables of the  system of two particles  partake in  correlations  that are proper of their quantum state and that give rise to  the very characteristic  property of \textit{entanglement}  (for a review, see~\cite{Horodecki:2009zz}). Because they are entangled, the two particles share properties---most notably, their spin correlations---that can only be discussed in the system as a whole.

Though a preeminent feature in atomic physics, quantum correlations among the components of a system has been somewhat played down in quantum field theory  because of the fixed momentum representation of the states in Fock space  and the commuting nature of the momentum and occupation number variables. 
Nevertheless, the quantum nature of the particles produced in high-energy collisions is there and their study could lead to new insights into their interaction.

Probing entanglement at collider was  first proposed in~\cite{Tornqvist:1980af,Abel:1992kz}. Tests  in the high-energy  regime of particle physics have been suggested  by means of neutral  meson physics~\cite{Benatti2,Banerjee:2016}, Positronium~\cite{Acin:2000cs}, Charmonium decays~\cite{Baranov:2008zzb}  and neutrino oscillations~\cite{Banerjee:2015mha}.
A discussion  of some of these issues also appears in~\cite{Yongram:2013soa,Cervera-Lierta:2017tdt}. 

The interest has been revived recently after entanglement has been shown \cite{Afik:2020onf} to be present in top-quark pair production at the LHC and it was  argued \cite{Fabbrichesi:2021npl} that  Bell inequalities violation is experimentally accessible in the same system. Following this lead, there has been more work about top-quark  production~\cite{Severi:2021cnj,Larkoski:2022lmv,Aguilar-Saavedra:2022uye,Afik:2022kwm}, hyperons~\cite{Gong:2021bcp} and gauge bosons  from Higgs boson decay~\cite{Barr:2021zcp,Barr:2022wyq}. 
For all these particles, it is possible to study entanglement and verify the violation of Bell inequalities.  

The same framework suggests the possibility of studying what happens when the SM amplitudes are modified by introducing new physics  beyond the SM. Because of its being so very fragile, entanglement provides a very sensitive probe to possible new physics in those cases where the SM states are produced in an entangled state which the new physics tends to lessen, modify or  brake altogether. 
In general, by adding extra terms to the SM interactions, the entanglement of the states is modified and the amount of change is a sensitive function of the  new physics present---which allows for direct constraints to be set. 

A first study of the impact of effective operators of dimension six---within the SM Effective Field Theory (SMEFT)---has  been presented in~\cite{Aoude:2022imd} for the entanglement of the spins of top-quark pairs. In this work we want to provide a comprehensive discussion for the case of bipartite, two-qubit like systems available at colliders and explore to what extent entanglement can provide a new tool in the search of physics beyond the SM.

We find that entanglement provides a novel set of observables that could lead to improved constraints  with respect to  those extracted from total cross sections or classical correlations. 
Quantum correlations can readily be studied in a bipartite system
made of either two spin-1/2 particles or two massless spin-1 (photons)~\cite{Clauser:1978ng}. Polarizations are measured at  colliders only for heavy fermions, the decays of which act as their own polarimeters; for this reason, we  study in detail the system of top-quark and $\tau$-lepton pairs produced at colliders. For the $\tau$-lepton pairs we also discuss the  case of their coming from the decay of the Higgs boson. 
In addition, we  include the case of the decay of the Higgs boson into two photons  because it can be modeled as a two-qubit system too and the framework is analogous to that of two fermions. We model the new physics by considering the effect of  representative operators not present within the SM and use  the entanglement observables to constrain the size of their contribution. 

For all these four systems we also check the Bell inequalities and find that they are maximally violated in the case of the Higgs boson decays.

New analytical results for the polarization matrix of the  Drell-Yan processes $q\bar q \to \tau \bar{\tau}$ in the SM and beyond, as well as the Higgs decay into $\tau^+ \tau^-$ and into two photons, are provided in the Appendix, which also includes known analytical results for the top-quark pair production  $q\bar q \to t \bar{t}$.

\section{Methods}

The quantum state of a bipartite system that can be modeled as a two-qubit pair, can be represented by the following
Hermitian, normalized, $4 \times 4$ density matrix:
%
%\begin{widetext}
\be
\rho  = \frac{1}{4}\Big[ \mathbb{1} \otimes \mathbb{1} + \sum_i B_i^+ (\sigma_i\otimes \mathbb{1} )
+ \sum_j B_j^-(\mathbb{1} \otimes \sigma_j)  
+ \sum_{ij} C_{ij} (\sigma_i\otimes\sigma_j) \Big]\ ,
\label{rho}
\ee
%\end{widetext}
%
where $\sigma_i$ are Pauli matrices, $\mathbb{1} $ is the unit $2\times 2$ matrix,
while the sums of the indices $i$, $j$ run over the labels 
representing any orthonormal reference frame in three-dimensions.

The real coefficients $B_i^+={\rm Tr}[\rho\, (\sigma_i\otimes \mathbb{1})]$ and
$B_j^-={\rm Tr}[\rho\, (\mathbb{1} \otimes\sigma_j)]$ represent the polarization
of the two qubits, while the real matrix $C_{ij}={\rm Tr}[\rho\, (\sigma_i\otimes\sigma_j)]$
gives their correlations. 
In the case of the particle pair system, $B_i^\pm$ and $C_{ij}$ are functions of the parameters
describing the kinematics of the  pair production. In addition, these coefficients need to satisfy further
constraints coming from the positivity request, $\rho\geq 0$, that any density matrix should fulfill; these extra
conditions are in general non-trivial, as they originate from requiring all principal minors of the matrix $\rho$ to be non-negative.

The two-qubit state $\rho$ is separable if it can be expressed as a convex combination of two-qubit product states:
\be
\rho  = \sum_{ij} p_{ij}\,  \rho^{(1)}_i\otimes \rho^{(2)}_j\ ,\quad \text{with} \quad p_{ij}>0 \quad \text{and} \quad \sum_{ij} p_{ij}=1\, , \label{separable}
\ee
where $\rho^{(1)}_i$ and $\rho^{(2)}_j$ are single-qubit density matrices. All states $\rho$ that can not be written in the form of \eq{separable}
are called entangled and exhibit quantum correlations. 

Correlations along only one direction in a given frame of reference can only probe classical properties as in the case of  angular momentum conservation---the correlation in this case being that if, say, spin up for one particle  is measured in the direction $z$, necessarily spin down will be measured in the same direction for the other particle, assuming the initial state has spin zero. It is only by the simultaneous measurement of correlations along more axes (or different bases) that we can probe quantum correlations, in particular, by  measuring  non-commuting quantities for the two particles, as in the case of the spin along the $z$ direction for the first particle and the spin along the $x$ direction for the second one. This is the reason why  the full matrix $C_{ij}$ is required in the study of entanglement.

Quantifying the entanglement content of a quantum state is in general
a hard problem, but for bipartite systems made of two qubits, an easily computable measure is available, 
the {\it concurrence}~\cite{Bennett:1996gf}.
It is constructed using the auxiliary $4\times 4$ matrix
\be
R=\rho \,  (\sigma_2 \otimes \sigma_2) \, \rho^* \, (\sigma_2 \otimes \sigma_2)\, , \label{rR}
\ee
where $\rho^*$ denotes the matrix with complex conjugated entries. Although non-Hermitian, the matrix $R$
possesses non-negative eigenvalues; denoting with $\lambda_i$, $i=1,2,3,4$, their square roots
and assuming $\lambda_1$ to be the largest,
the concurrence of the state $\rho$ is defined as
\be
{\cal C}[\rho] = \max \big( 0, \lambda_1-\lambda_2-\lambda_3-\lambda_4 \big)\ .
\label{concurrence}
\ee
Concurrence vanishes for separable states, like those defined in (\ref{separable}), reaching its maximum value 1 when $\rho$ is
a projection on a pure, maximally entangled state.

The presence of entanglement in a quantum system, that is of correlations among its constituents not
accounted for by classical physics, can lead to the violation of suitable constraints, the so-called
Bell inequalities, that are instead satisfied by certain local, stochastic completions of quantum mechanics~\cite{Bell:1964kc,Bell2,Clauser:1978ng}.
In the case of a two-qubit system in the state (\ref{rho}), as a pair of spin-1/2 particles, 
a very useful test is provided by the
the following inequality involving only the correlation matrix $C$~\cite{Horodecki2}:
\be
\Big|{\hat n}_1\cdot C \cdot \big({\hat n}_2 - {\hat n}_4 \big) +
{\hat n}_3\cdot C \cdot \big({\hat n}_2 + {\hat n}_4 \big)\Big|\leq 2\ ,
\label{CHSH}
\ee
where ${\hat n}_1$, ${\hat n}_2$, ${\hat n}_3$ and ${\hat n}_4$
are four different three-dimensional unit vectors determining four spatial directions,
along which the spins of the two particles can be measured. In order to test this (generalized) Bell inequality
one needs to maximize the left-hand side of (\ref{CHSH}) by a suitable choice of the four spatial directions.
In practice, this maximization procedure can be overcome by looking at the eigenvalues $m_1$, $m_2$, $m_3$, of the
symmetric, non-negative, $3\times 3$ matrix $M=C^T C$, where $C^T$ is the transpose of $C$, that can be 
ordered in decreasing magnitude $m_1\geq m_2\geq m_3$.
At this purpose it is convenient to introduce the operator $\mathfrak{m}_{12}[C]$ defined as
\be
\mathfrak{m}_{12}[C]\equiv m_1 + m_2\, .
\label{C12}
\ee
 As proven in \cite{Horodecki2}, given a two-qubit state $\rho$ as in
     (\ref{rho}), with a correlation matrix $C$ satisfying the condition
\be
\mathfrak{m}_{12}[C] >1\, .
\label{inequality-test}
\ee
then there surely are choices for the vectors  ${\hat n}_1$, ${\hat n}_2$, ${\hat n}_3$, ${\hat n}_4$ 
for which the left-hand side of (\ref{CHSH}) is larger than 2;
in other words, the two-qubit state (\ref{rho}) violates (\ref{CHSH}) if and only if the sum of the two largest eigenvalues of $M$ is strictly larger than 1.

In the following we  concentrate on the two observables:
\be
{\cal C}[\rho] \quad \text{and} \quad \mathfrak{m}_{12}[C]\ . 
\label{obs}
\ee
We use the first  to constrain possible new physics extension of the SM by studying spin correlations 
in top-quark and $\tau$-lepton pairs produced at colliders, the second to check violations of the Bell inequalities in the same pair systems as well as in the decay of the Higgs boson into  $\tau$-lepton pairs and two photons. 

Although other entanglement witnesses have been considered in high-energy physics \cite{Afik:2020onf,Aoude:2022imd},
we stress that the concurrence ${\cal C}[\rho]$ directly and fully quantifies the entanglement content of the state $\rho$, it can be easily
computed and readily measured in experiments.
The observable $\mathfrak{m}_{12}[C]$ witnesses Bell non-locality and thus it is perfectly suited to test the Bell inequality
(\ref{CHSH}); indeed, as already remarked, it automatically selects the best choice of the four units vectors ${\hat n}_i$ that maximizes the left-hand side of (\ref{CHSH}) and thus its violation.

\subsubsection{Kinematics and projector operators}

We consider  the cross section for the process in which two parton quarks go into two final fermions
\be
q(q_1) + \bar q(q_2) \to f(k_1) + \bar f (k_2)\, .
\ee
The momenta $k_1$ and $k_2$ of the final  fermion and anti-fermion, and  $q_1$ and $q_2$ of the entering quark and anti-quark, respectively,  can be written  in the center-of-mass (CM) system as
\bea
k_1 &= & \left( \frac{m_f}{\sqrt{1-\beta_f^2}},\, \frac{m_f \beta_f \sin \Theta}{\sqrt{1-\beta_f^2}},\, 0, \, \frac{m_f \beta_f \cos \Theta}{\sqrt{1-\beta_f^2}}\right) \nn \\
k_2&= & \left( \frac{m_f}{\sqrt{1-\beta_f^2}},\,- \frac{m_f \beta_f \sin \Theta}{\sqrt{1-\beta_f^2}},\, 0, \, -\frac{ m_f \beta_f \cos \Theta}{\sqrt{1-\beta_f^2}}\right)  \nn \\
q_1 &= & \left( \frac{m_f}{\sqrt{1-\beta_f^2}},\, 0,\, 0, \, \frac{m_f}{\sqrt{1-\beta_f^2}}\right) \nn \\
q_2&= & \left( \frac{m_f}{\sqrt{1-\beta_f^2}},\,0,\, 0, \, -\frac{ m_f }{\sqrt{1-\beta_f^2}}\right) \, , \label{momenta}
\eea
where $m_f$ is the mass of the final fermions and 
\be
\beta_f = \sqrt{1 -  4 \frac{m_f^2}{m_{f\bar f}^2}}\, , \label{beta_f}
\ee
where $m_{f\bar f}$ is the fermion-pair invariant mass, with $\Theta$ the angle between the initial and final fermion momenta in the CM frame.

Spin correlations, as embodied by the coefficients $C_{ij}$ in \eq{rho}, are extracted from the cross section
 by taking the product  of the polarizations of the final fermions. This is done by means of the usual projectors over definite polarizations (for a fermion  with momentum $p$, mass $m_f$ and polarization vector $\zeta$): 
\be
 u(p, \zeta)  \otimes \bar u(p, \zeta) = \frac{1}{2} (\slashed{p} + m_f) ( 1 - \gamma_5 \slashed \zeta) 
\quad \text{and} \quad v(p, \zeta)  \otimes\bar v(p,\zeta) = \frac{1}{2} (\slashed{p} - m_f) ( 1 - \gamma_5 \slashed \zeta)  \, , \label{fproj}
\ee
inserted in the square of the modulus of the amplitude. The coefficients $B_i^\pm$ in \eq{rho} are instead obtained by keeping only one of the two particle polarizations.

We adopt the orthonormal basis introduced in~\cite{Bernreuther:2013aga} in order to describe the spin correlations thus obtained.
Let $\hat{p}$ the unit vector along one of the proton beam directions in the laboratory frame
and denote with $\hat{k}$ the direction of flight of the final fermion in the fermion pair CM frame;
then, a convenient reference frame is defined by the three unit vectors:
\be
\hat{r}=\dfrac{1}{r}(\hat{p}-y\hat{k})\ ,\qquad  \hat{n}= \dfrac{1}{r}(\hat{p}\times \hat{k})\ ,
\label{eq:axes}
\ee
where
\be
y=\hat{p}\cdot \hat{k}=\cos\Theta\ ,\qquad r=\sqrt{1-y^2}\ ,
\ee
with $\Theta$ being the  scattering angle. Notice, that at partonic level  the angle $\Theta$ is defined, according to the momenta in \eq{momenta}, as the angle between  initial quarks and final fermion.

The elements $C_{ij}$ of the correlation matrix in \eq{rho}  are obtained on the various components of the chosen basis by means of the polarizations vectors:
\bea
\zeta_1^k &= & \left( \frac{\beta_f}{\sqrt{1-\beta_f^2}},\, \frac{\sin \Theta}{\sqrt{1-\beta_f^2}},\, 0, \, \frac{\cos \Theta}{\sqrt{1-\beta_f^2}}\right) \nn\\
\zeta_2^k &=&  \left(- \frac{\beta_f}{\sqrt{1-\beta_f^2}},\, \frac{\sin \Theta}{\sqrt{1-\beta_f^2}},\, 0, \, \frac{\cos \Theta}{\sqrt{1-\beta_f^2}}\right) \nn \\
\zeta_1^r &= & \zeta_2^r = \left(0,\, -\cos\Theta ,\, 0, \, \sin\Theta \right) \nn \\
\zeta_1^n &=&  \zeta_2^n= \left(0,\, 0 ,\, 1, \, 0 \right) \label{spins}
\eea
where the indices 1 and 2 stand for the final fermion and anti-fermion.

 The differential cross section for pair $f \bar f$ production is given by
 \be
 \frac{\di \sigma^{f \bar f}}{\di \Omega} = \frac{\beta_f\,  |\overline{\cal M}|^2}{64 \pi^2 \, m_{f\bar f}^2} \, ,
 \ee
where $\cal M$ is the amplitude for the production and $\beta_f$ is defined in \eq{beta_f}.

The parton level differential cross section for the two partons---with fraction $x_1$ and $x_2$ of the available momentum and momentum distribution (PDF) $q(x)$ and $\bar q(x)$---is given by
\be
\frac{\di \sigma^{f \bar f}}{\di \Omega \,\di m_{f\bar f}} =  2 \int \frac{\di \sigma^{f \bar f}}{\di \Omega} q(x_1) \bar q (x_2)  \delta( m^2_{f\bar f} - x_1 x_2 s)\, \di x_1 \di x_2\, , \label{x-sec}
\ee
for   CM energy $\sqrt{s}$. 
The cross section in \eq{x-sec} can be  re-written in terms of the parton luminosity function 
\be
L^{q\bar q}(\tau) =\frac{4 \tau}{\sqrt{s}} \int_\tau^{1/\tau} \frac{\di z}{z} q_{q} (\tau z) q_{\bar q} \left( \frac{\tau}{z}\right) \label{pf}
\ee
by the change of variables $x_1=\tau/z$ and $x_2 = \tau z$, with $m^2_{f \bar f} = \tau^2 s$. In \eq{pf} there is an overall factor 2 due to the symmetrization between quark and anti-quark. We use the parton luminosity function in \eq{pf}, as well as the corresponding one for  the case of having gluons as the partons,  in what follows. The differential cross section in \eq{x-sec} can thus be written as
\be
\frac{\di \sigma^{f \bar f}}{\di \Omega \,\di m_{f\bar f}} = \frac{\beta_f\,  |\overline{\cal M}|^2}{64 \pi^2 \, m_{f\bar f}^2} L^{q\bar q}(\tau) \, ,
\ee
where $ L^{q\bar q}(\tau)$ stands for  the  luminosity functions of the corresponding quark or gluon partons.

In the case of the Higgs boson $H$ decay 
\be
H\to f(k_1) \bar{f}(k_2)\, ,
\ee
the same four-vectors in \eq{momenta} can be used by imposing $m_{f\bar f}=m_h$ and putting $q_1$ and $q_2$ at rest and equal; the vectors $k_1$ and $k_2$ are back-to-back with no scattering angle dependence. The polarization of the final states follows the same structure as given in \eq{fproj}, with polarization vectors as in \eq{spins}. The corresponding $C_{ij}$ elements are obtained in the same way as above, by projecting on the various components of the chosen basis for the polarization vectors in \eq{spins}.

Regarding the Higgs boson decay into two photons
\be
H\to \gamma(k_1)\,  \gamma(k_2)\, ,
\label{Hgg}
\ee
the projection in the product of the associated two photon polarizations $e_\mu^{\lambda_1}(k_1)$ and $e_\mu^{\lambda_2}(k_2)$ (with both $\lambda_1$ and $\lambda_2$ indices taking values  1 and 2) can be obtained in similar fashion by expressing the corresponding density matrix $\rho_{\mu\nu}$ for a photon of generic momentum $k$ and polarization $e_\mu^{\lambda_1}(k)$ as a function of the Stokes parameters $\xi_i$ :
\bea
\rho_{\mu\nu}(\vec{\xi})&=&% \sum_{\lambda,\lambda^{\prime}}
 e_\mu^{\lambda}(k) e_\nu^{\lambda^{\prime}\ast}(k)\,=\,
\frac{1}{2} \hat{e}^{T}_\mu \Big( \mathbb{1}  + \vec \xi \cdot \vec \sigma \Big) \hat{e}_\nu  \nn \\
 & =& \frac{1}{2} 
 \Big( e_\mu^{(1)} e_\nu^{(1)} + e_\mu^{(2)} e_\nu^{(2)} \Big) +\frac{\xi_1}{2} 
 \Big( e_\mu^{(1)} e_\nu^{(2)} + e_\mu^{(2)} e_\nu^{(1)} \Big)\nn \\
 & &  - \frac{i\xi_2}{2} 
 \Big( e_\mu^{(1)} e_\nu^{(2)} - e_\mu^{(2)} e_\nu^{(1)} \Big) 
 + \frac{\xi_3}{2} 
 \Big( e_\mu^{(1)} e_\nu^{(1)} - e_\mu^{(2)} e_\nu^{(2)} \Big)\, ,
 \label{densitymat}
 \eea
 where the compact vectorial notation $\hat{e}_{\mu}\equiv (e_{\mu}^{(1)},e_{\mu}^{(2)})$ is adopted, with $\hat{e}^T_{\mu}$ standing for the transpose, $k$ the photon 4-momentum, and $\sigma_i$ the Pauli matrices; the four-vectors $e^{(\lambda)}_\mu$ provide a basis for the linear polarizations, they are two ortho-normal four-vectors orthogonal to the momentum: $e^{(\lambda)}\cdot e^{(\lambda^\prime)}=-\delta^{\lambda\, \lambda^\prime}$,  $e^{(\lambda)}_\mu \cdot k=0$. 
 To lighten the notation, we removed the momentum dependence inside the polarization basis.

 For the two photon system, the product $\rho(\vec{\xi}^{(1)})\rho(\vec{\xi}^{(2)})$  of the two density matrix enters, with $\vec{\xi}^{(1)},\vec{\xi}^{(2)}$ the corresponding Stokes parameters of the two photons.  In this case the correlation matrix $C_{ij}$ for the two photon system is expressed on the basis of the Stokes parameters and it can be simply extracted by selecting the terms proportional to $\xi_i^{(1)}\xi_j^{(2)}$ in the expression for the polarized amplitude square $|{\cal M}^2|$ of the process, and dividing them for the unpolarized $|\bar{{\cal M}}^2|$ contribution. Analogously, for the $B_i^+$ or $B_i^-$ terms in
 \eq{rho}, which can be extracted taking the linear terms proportional to the corresponding Stokes parameters $\xi_i^{(1)}$ or $\xi_i^{(2)}$ respectively, in the expression of $|{\cal M}^2|$ and normalizing them by  $|\bar{{\cal M}}^2|$.

\subsubsection{Uncertainty estimates}

The setting of bounds on physics beyond the SM is all about the knowing the uncertainty of our limits. To estimate the intrinsic uncertainty in the determination of observables in \eq{obs} for the various processes we are going to consider, we run 1000 pseudo-experiments according to the probability distribution of the observables themselves.

For example, in comparing SM and new physics,
from the    distributions of the observable value in the scattering angle and the invariant mass obtained for the two cases, we can obtain the  significance with which we can separate the two. To quantify this difference in terms of statistical significance, we compute the p-value 
 of the new physics  distribution by integrating the SM distribution  from the mean value  to $-\infty$.  

The  significance is defined as ${\cal Z} = \Phi^{-1}(1-p)$
where
\be
\Phi(x) = \frac{1}{2}\, \left[ 1 + \text{erf} \left(\frac{ x}{\sqrt{2}} \right) \right]\,.
\ee
The value of  ${\cal Z}$ assigns a statistical significance to the separation between the two distributions. We can take $ {\cal Z}$ as the number of standard deviations $\sigma$, in the approximation in which the distribution is assumed to be Gaussian, and translates the number of standard deviations into a confidence level (CL). 

In the case of the Higgs boson decays, since there is no kinematical variation, we simply draw a Gaussian dispersion with $\sigma = 1/\sqrt{N}$, where $N$ is taken from the corresponding number of events generated by the production cross sections and branching ratios multiplied for the benchmark luminosities. In those cases for which this number is very large, we simply take 1000 events as an illustrative example.

As we  comment further for the specific examples, all these estimates of uncertainty are limited by being performed at the level of  primary parton production. They do not take into account the additional uncertainty coming from the extraction of the entanglement observables from the actual data---which are the angular distributions of the final states originating from the decays of the top quarks and  $\tau$-leptons. As shown in \cite{Fabbrichesi:2021npl} for the case of the top quark pairs, we do expect a substantial additional uncertainty from systematic errors in the reconstruction. In addition, there may also be other confounding contributions from possible background events.

\section{Results: Top quark pairs} 

Top quark pairs are routinely produced at the LHC
and the  spin correlations among quark pairs 
has  been shown~\cite{Mahlon:1995zn,Bernreuther:2001rq,Bernreuther:2010ny,Mahlon:2010gw,Bernreuther:2015yna} to be a powerful tool  in the physical analysis---limited aspects of which have been already studied by the experimental collaborations at the LHC on  data at 7 \cite{ATLAS:2012ao}, 8~\cite{Aad:2014mfk} and 13 TeV~\cite{CMS:2018jcg} of CM energy.

In this section we study the operators in \eq{obs} in the case of top-quark pairs and use them to constrain new physics and check the Bell inequalities.

\subsection{Entanglement in $t\bar t$ production}

Before plunging into the actual analysis of  spin correlations in the top-quark pair production, it is useful
to discuss qualitatively what the SM predicts for the entanglement content of the $t \bar t$-spin state as
described by a density matrix $\rho_{t\bar t}$ as in (\ref{rho}).

The dependence of the entries of the matrix (\ref{rho})
on the kinematic variables $\Theta$, the scattering angle,  and $\beta_t $, as defined in \eq{beta_f}, is in general rather involved but
it simplifies  at $\Theta=\pi/2$ for which the top-quark pair is  transversally produced and the entanglement is  maximal. In this limit, we choose the three vectors
$\{\hat{r},\, \hat{k}, \,\hat{n}\}$ to point in the $\{\hat{x},\hat{y},\hat{z}\}$ directions.
In this frame, let us denote by $|0\rangle$ and $|1\rangle$ the eigenvectors of the Pauli matrix $\sigma_z$
with eigenvalues $-1$ and $+1$, respectively; similarly, let $|\small{-}\rangle$ and $|\small{+}\rangle$ be the analogous
eigenvectors of $\sigma_x$ and $|\text{\small L}\rangle$ and $|\text{\small R}\rangle$ those of $\sigma_y$.

Possible quark pair spin density matrices can  be both projectors on pure, maximally entangled Bell states,
\be
\rho^{(\pm)}= |\psi^{(\pm)}\rangle\langle\psi^{(\pm)}|\ ,\qquad 
|\psi^{(\pm)}\rangle=\frac{1}{\sqrt{2}}\big( |01\rangle \pm |10\rangle \big)\ ,
\label{Bell-states}
\ee
and mixed, unentangled states,
\begin{eqnarray}
&&\rho^{(1)}_{\rm mix}=\frac{1}{2}\Big( |{\small ++}\rangle\langle {\small ++}| + |\small{--}\rangle\langle {\small --}| \Big)\ ,\\
\label{rho-mix-2}
&&\rho^{(2)}_{\rm mix}=\frac{1}{2}\Big( |\text{\small LR}\rangle\langle \text{\small LR}| + |\text{\small RL}\rangle\langle \text{\small RL}| \Big)\ ,\\
&&\rho^{(3)}_{\rm mix}=\frac{1}{2}\Big( | 01 \rangle\langle 01| + |10\rangle\langle 10| \Big)\ .\\
\nonumber
\end{eqnarray}

Let us  treat separately the quark-antiquark $q \bar{q}$ and 
gluon-gluon $gg$ production channels.
For the $q \bar{q}$ production channel, using the explicit expression   collected in the Appendix for the correlation coefficients $C_{ij}$, one obtains that 
the $t \bar{t}$ spin density matrix can be expressed as the following convex combination:
\be
\rho_{t\bar t}^{(q\bar{q})}= \lambda \rho^{(+)} + (1-\lambda)\rho^{(1)}_{\rm mix}\ ,\quad \text{with} \quad \lambda=\frac{\beta_t^2}{2-\beta_t^2}\in [0,1]\ , 
\label{rho-qq}
\ee
so that at high transverse momentum, $\beta_t \rightarrow 1$, the spins of the $t\bar{t}$ pair 
tend to be generated in a maximally entangled state; this quantum correlation is however progressively
diluted for $\beta_t < 1$, vanishing at threshold, $\beta_t=0$, as the two spin state becomes a
totally mixed, separable state.

The situation is different for the $gg$ production channel, as both at threshold and at high momentum
the $t \bar{t}$ spins result maximally entangled, with $\rho_{t\bar t}^{(gg)}=\rho^{(+)}$ for $\beta_t \rightarrow 1$
and $\rho_{t\bar t}^{(gg)}=\rho^{(-)}$ when $\beta_t=\, 0$. For intermediate values of $\beta_t$, the situation becomes more involved,
and the two-spin density matrix can be expressed as the following convex combination:
\be
\rho_{t\bar t}^{(gg)}= a\rho^{(+)} + b \rho^{(-)} + c \rho^{(1)}_{\rm mix} + d \rho^{(2)}_{\rm mix}\ ,
\ee
with non-negative coefficients
\be
a=\frac{\beta_t^4}{1+2\beta_t^2-2\beta_t^4}\ ,\quad
b=\frac{(1-\beta_t^2)^2}{1+2\beta_t^2-2\beta_t^4}\ ,\quad
c=d=\frac{2\beta_t^2\big(1-\beta_t^2\big)}{1+2\beta_t^2-2\beta_t^4}\ ,\quad
\ee
so that $a+b+c+d=1$, while entanglement is less than maximal.

Putting together the $q \bar{q}$- and $gg$-contributions, as discussed below,  leads to more
mixing and therefore in general to additional loss of quantum correlations. Nevertheless, this preliminary
analysis already indicates that in order to get the larger $t \bar t$-pair spin entanglement 
one has to look at the kinematic region of high energy and large scattering angle.

\subsection{ Computing the observables}

We compute all the entries of the correlation matrix $C_{ij}$  for the process
\be
p+ p \to t+\bar t  \, .
\label{eq:process}
\ee
with the unpolarized cross section given by
\be
\frac{\di \sigma}{\di \Omega\, \di m_{t\bar t}} = \frac{\alpha_s^2 \beta_t}{64 \pi^2  m_{t\bar t}^2} \Big\{ L^{gg} (\tau) \, \tilde A^{gg}[m_{t\bar t},\, \Theta]+L^{qq} (\tau)\, \tilde A^{qq}[m_{t\bar t},\, \Theta]  \Big\} \label{x-sec-tt}
\ee
where $L^{gg,qq}(\tau)$ are the parton luminosity functions defined in \eq{pf} of section II,  $\tau = m_{t\bar t}/\sqrt{s}$ and  $\alpha_s=g^2/4 \pi$. The explicit expressions for $\tilde A^{gg}$ and  $\tilde A^{qq}$  are given in the Appendix. 

\begin{figure}[h!]
\begin{center}
\includegraphics[width=5in]{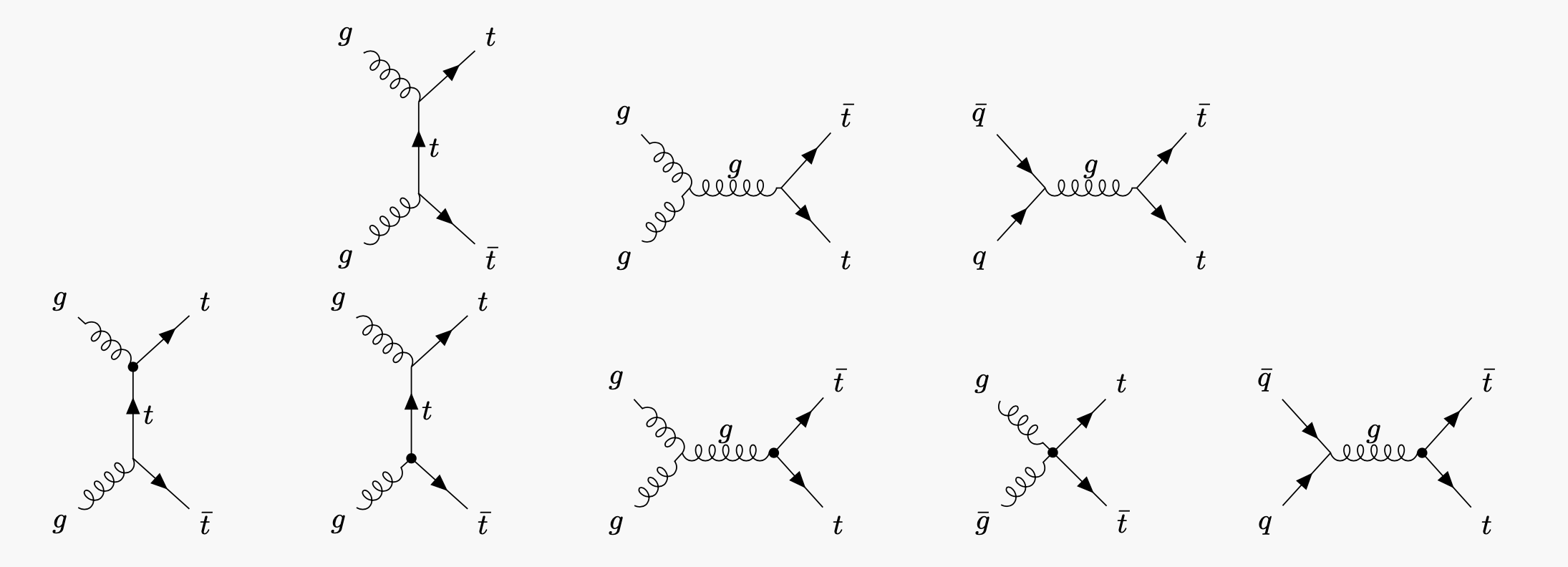}
\caption{\small \label{fig:ttbar} Feynman diagrams for $t\bar t$ production. The dot stands for the magnetic dipole vertex (see \eq{dipole}).}
\end{center}
\end{figure}

The combination of the two channels (see Fig.~\ref{fig:ttbar}) $g+g\to t +\bar t$ and $q + \bar q \to t +\bar t$ in \eq{x-sec-tt} is weighted by the respective  parton luminosity functions
\be
L^{gg} (\tau)= \frac{2 \tau}{\sqrt{s}} \int_\tau^{1/\tau} \frac{\di z}{z} q_{g} (\tau z) q_{g} \left( \frac{\tau}{z}\right)\quad \text{and}\quad
L^{qq} (\tau)= \sum_{q=u,d,s}\frac{4 \tau}{\sqrt{s}} \int_\tau^{1/\tau} \frac{\di z}{z} q_{q} (\tau z) q_{\bar q} \left( \frac{\tau}{z}\right)\, ,
\ee
where the functions $q_j(x)$ are the PDFs. Their numerical values are those provided by  a recent sets ({\sc PDF4LHC21})~\cite{PDF4LHCWorkingGroup:2022cjn} for $\sqrt{s}=13$ TeV and factorization scale $q_0=m_{t\bar t}$ (we have used for all our results the subset 40).

The correlation coefficients are given as
\be
C_{ij} [m_{t\bar t},\, \Theta]= \frac{L^{gg} (\tau)\, \tilde C_{ij}^{gg}[m_{t\bar t},\, \Theta]+L^{qq} (\tau)\, \tilde C_{ij}^{qq}[m_{t\bar t},\, \Theta]} {L^{gg}(\tau) \, \tilde A^{gg}[m_{t\bar t},\, \Theta]+L^{qq} (\tau)\, \tilde A^{qq}[m_{t\bar t},\, \Theta]} \, .\label{pdf}
\ee
The explicit expression for the coefficient $\tilde C_{ij}^{gg}$ and $\tilde C_{ij}^{qq}$ for the SM as well as for the new physics are collected in the Appendix.

The expression in \eq{pdf} must be expanded in the case of new physics by retaining terms linear in the new physics coefficient.
As already remarked, the combination of different  terms in the cross section implies a decrease in entanglement. 
For this reason, the parton luminosities play an important role. 

%%%%%%%%%%%%%%%%%%%%%%%%%%%%%%%%%%
%%%%%%%%%%%%%%%%%%%%%%%%%%%%%%%%%%%%%%%%%%%%%

%%%%%%%%%%%%%%
 \begin{figure}[h!]
\begin{center}
\includegraphics[width=3.5in]{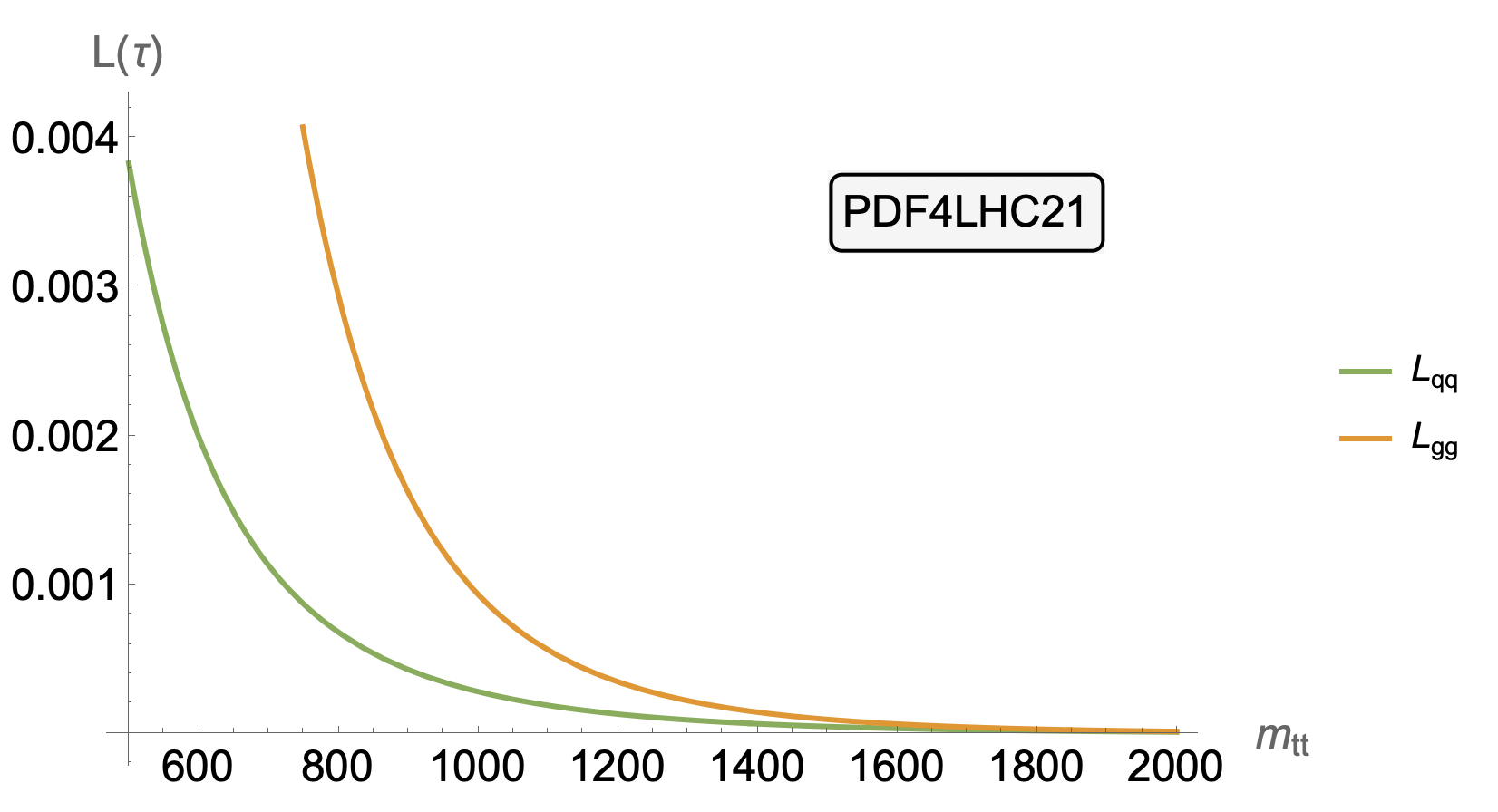}
\caption{\small  Parton luminosities: The gluon luminosity is about 9 times larger than the quark luminosity at threshold, then decreasing to being about 30\% larger around 1 TeV.
\label{fig:pdftop} 
}
\end{center}
\end{figure}
%%%%%%%%%%%%%%%%%%%%%% 

\subsection{Bell inequalities}

Let us first discuss the violation of Bell inequalities
coming from the  entanglement of the top-quark pair. This violation has been already discussed in \cite{Fabbrichesi:2021npl} by means of   a  numerical simulation of the data of run2 at the LHC and finding an estimate of  $\mathfrak{m}_{12}[C]>1$ with a CL 95\%.  The estimate of the operator $\mathfrak{m}_{12}[C]$ requires a correction for the inherent bias in the numerical computation of the eigenvalues. The operator is a  consistent estimator and its use  in testing the violation of Bell inequalities,  as in \cite{Fabbrichesi:2021npl},  valid. Such an analysis provides, as already mentioned, a more realistic estimate of the uncertainty than that we are going to provide here which is only based on the primary particle production.

Here we compute the violation of the Bell inequalities directly from the analytical expression of $\mathfrak{m}_{12}[C]$ in \eq{C12}. Because the only off-diagonal term in the matrix 
\be
C=  \begin{pmatrix} C_{nn}&C_{nr}&C_{nk}\\
C_{rn}&C_{rr}&C_{rk} \\
C_{kn}&C_{kr}&C_{kk} \label{matrixC}
\end{pmatrix} \ ,
 \ee
is $C_{kr}$, its eigenvalues are given by
\be
 C_{nn}^2 , \quad
 \frac{1}{4} \Big[ C_{kk} +  C_{rr} + \sqrt{(C_{kk}^2 - C_{rr})^2+ 4 C_{kr}^2}  \Big]^2 , \quad
  \frac{1}{4} \Big[ C_{kk} +  C_{rr}- \sqrt{(C_{kk}^2 - C_{rr})^2+ 4 C_{kr}^2} \Big]^2 \,. \label{eigenC}
\ee
The sum of the two largest among the three eigenvalues in \eq{eigenC}  give us the value of the operator $\mathfrak{m}_{12}[C]$.

%%%%%%%%%%%%%%
 \begin{figure}[h!]
\begin{center}
\includegraphics[width=3.3in]{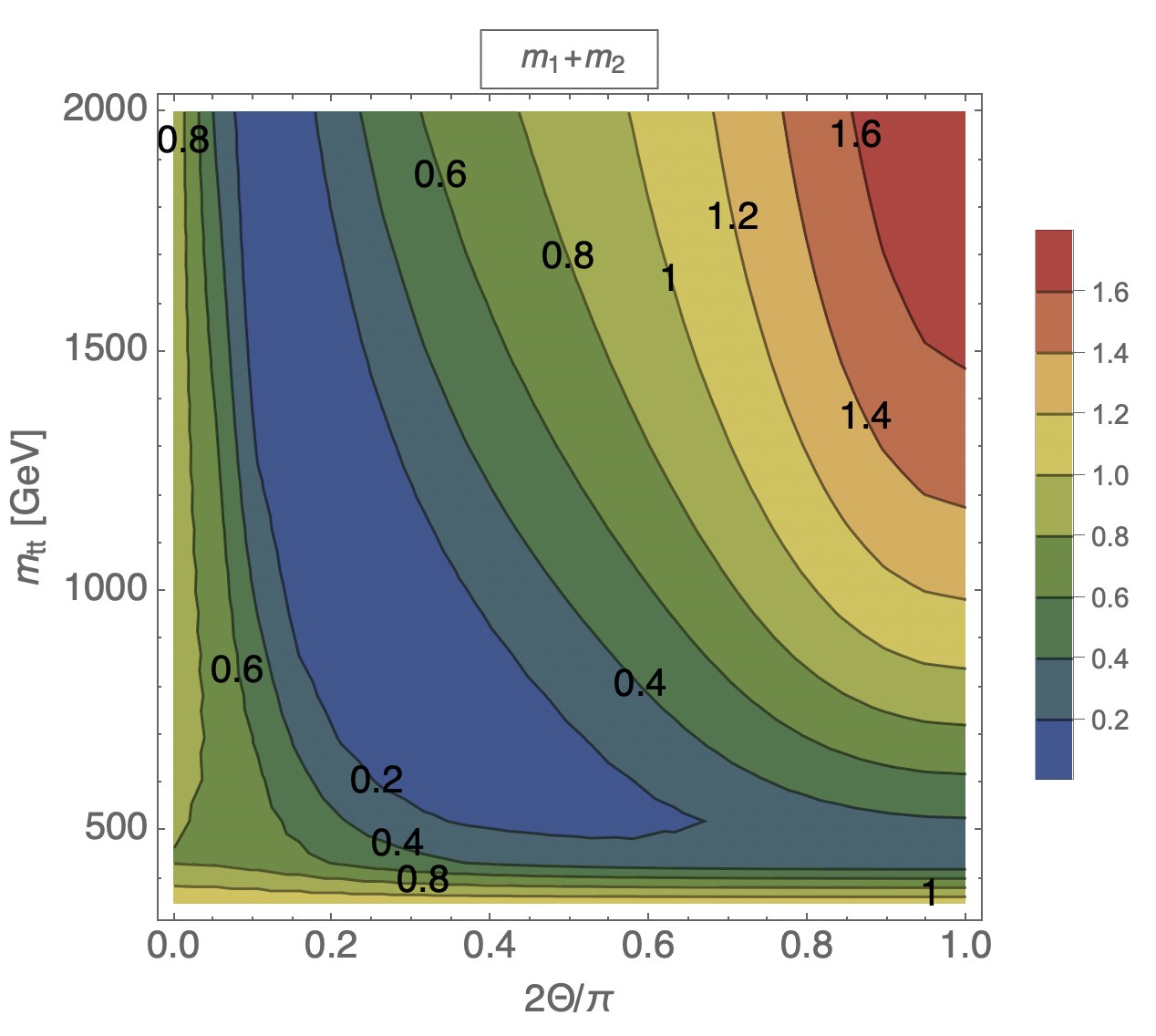}
\caption{\small The observable  $\mathfrak{m}_{12}[C]$ as a function of the kinematical variables $\Theta$ and $m_{t \bar t}$ across the entire available space. 
\label{fig:m1m2} 
}
\end{center}
\end{figure}
%%%%%%%%%%%%%%%%%%%%%% 

The values of the observable $\mathfrak{m}_{12}[C]$ across the entire kinematical space available are shown in the contour plot in Fig.~\ref{fig:m1m2}. In this and the following contour plots the values of the observable are symmetric for $1<2 \Theta/\pi<2$. The figure shows how the quantum entanglement increases as we consider larger scattering angles and, as expected from the qualitative discussion in section III.A, is maximal for $m_{t\bar t} > 900$. Therefore, we zoom in the kinematical window where the observable $\mathfrak{m}_{12}[C]$ is larger, namely for $m_{t\bar t} > 900$ GeV and $2 \Theta/\pi > 0.85$. The mean value of $\mathfrak{m}_{12}[C]$ in this bin is 1.44.

 We give in Table~\ref{tab:events_top} the number of events for the two benchmark cases of the run 2 LHC and the future Hi-Lumi.  Cross sections are computed by running {\sc MADGRAPH5}~\cite{Alwall:2014hca} at the LO and then correcting by the $\kappa$-factor given at the NNLO~\cite{Czakon:2020qbd}.
These are events in which the two tops decay into leptons ($\BR 2.3\%$)
%%%%%%%%%%%%
\begin{table}[h!]
\bc
\begin{tabular}{ccc}
%\toprule
&\hskip0.5cm (run 2) ${\color{oucrimsonred} {\cal L}=140\ \text{fb}^{-1}}$  \hskip0.5cm &  \hskip0.5cm (Hi-Lumi) ${\color{oucrimsonred} {\cal L}=3\ \text{ab}^{-1}}$  \hskip0.5cm \\[0.2cm]
\hline\\
 \underline{events} \hskip0.4cm  &\hskip0.4cm  $463$  \hskip0.4cm &\hskip0.4cm $9727$ \hskip0.4cm \\[0.4cm]
   \hline%
\end{tabular}
\caption{\label{tab:events_top}Number of expected events in the kinematical region $m_{tt}>900$ GeV and $0.85<x<1$.}
\ec
\end{table}
%%%%%%%%%%%%%%%%%%

To estimate the intrinsic uncertainty, we run 463 pseudo-experiments according the probability distribution of our observable. We find a significance of 55 for the hypothesis $\mathfrak{m}_{12}[C]>1$ already with the 463 events from run2 at the LHC.

A segmentation of the kinematical window into smaller bins (the values of which can then be collected in a $\chi^2$ test) improves the significance of the violation, as shown in~\cite{Fabbrichesi:2021npl}.   

%%%%%%%%%%%%%%
 \begin{figure}[h!]
\begin{center}
\includegraphics[width=1.62in]{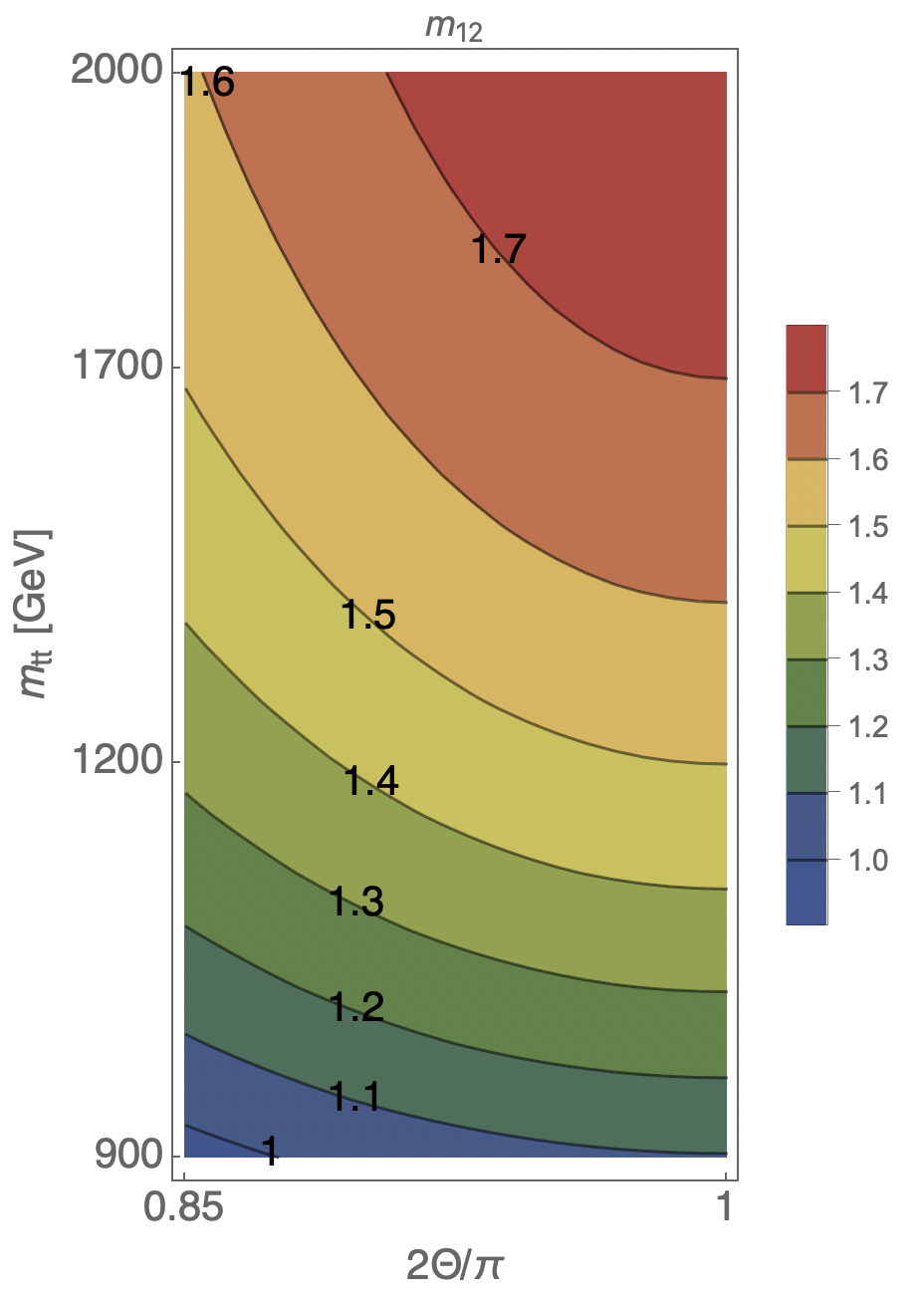}
\includegraphics[width=2.6in]{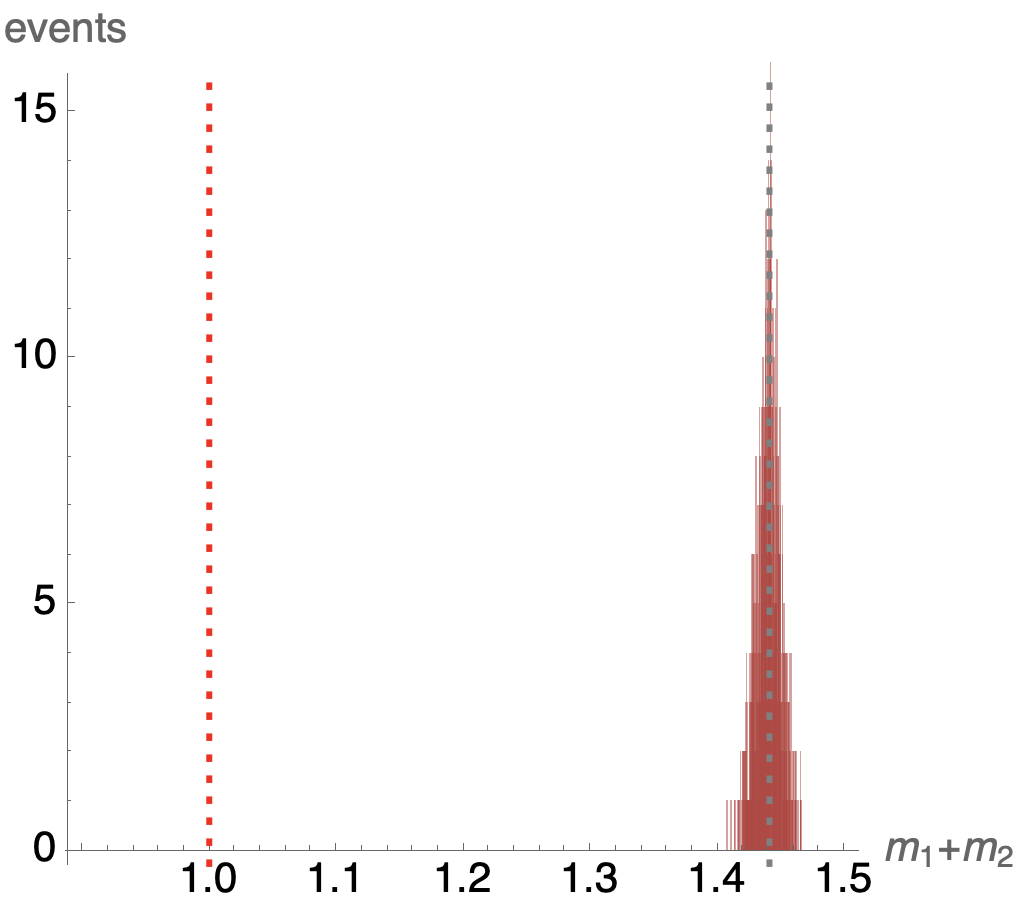}
\caption{\small \underline{On the left}: The observable  $\mathfrak{m}_{12}[C]$ in the kinematical window $m_{t\bar t} > 900$ GeV and $2 \Theta/\pi > 0.85$. \underline{On the right}: The statistical distribution for 463 events (with mean value $1.44$ and dispersion  $\sigma=0.008$) compared with the critical value 1 above which Bell inequalities are violated.
\label{fig:significance} 
}
\end{center}
\end{figure}
%%%%%%%%%%%%%%%%%%%%%% 

 The main source of theoretical uncertainty comes from 
  higher order QCD corrections to the LO values of the $C_{ij}$ matrix elements
  in \eq{matrixC}. Following the results in \cite{Bernreuther:2001rq,Czakon:2015owf}, where the NLO QCD corrections have been computed, we find that the error induced by these missing corrections to the largest eigenvalues in \eq{eigenC} is of the order of 8\%, which gives approximatively a 10\% uncertainty on the main
  entanglement observables in the relevant kinematic regions.
  Small sources of theoretical uncertainties come from the PDF and the top-quark mass, but these are negligible. By comparing results with two different set of PDF, we estimate the related uncertainty to be of the order of per mille. This is of the same order as the uncertainty induced by top-quark mass, obtained by varying its mass within the two standard deviation of its experimental value.

Concerning other sources of uncertainties, we stress here that the analysis is not affected by  possible backgrounds.  In the top-quark pair   production, the background 
consists of  extra contributions  to the $\ell^-\ell^+\nu\bar{\nu} b\bar{b}$ final state, which are negligible once the kinematic of the two on-shell top quarks is fully reconstructed. Further sources of  background include $t\bar{t}W$ or $t\bar{t}Z$  as well as di-boson events, and misidentification of leptons---which amount to a few percent of the total. Another source of  background is provided by the $Z$+jets events, whose numbers, after the cuts, is comparable to the other backgrounds aforementioned. 

 It goes without saying that this is an estimate that does not take into account the systematic errors inherent in the procedure of obtaining the observable from the actual data. Taking the result in \cite{Fabbrichesi:2021npl} as guidance, we must be ready to see the  significance  drop by about one order of magnitude.

\subsection{New physics: the magnetic dipole moment}

As a benchmark in searching for new physics, we consider the presence of a magnetic dipole operator in the coupling between the top quark and the gluons:
\be
{\cal L}_{\text{\tiny  dipole}}= - \mu \, \frac{g_s}{2 m_t} \, \bar t \, \sigma^{\mu \nu} \, T^a\, t \, G_{\mu\nu}^a\, . \label{dipole}
\ee

In comparing our results with those  following the conventions  of the SMEFT, with the magnetic dipole Lagrangian for the top quark expressed in a $SU(2)_L$ invariant way as
\be
{\cal L}_{\text{\tiny  dipole}}^{\prime}=\frac{c_{tG}}{\Lambda^2} \big( {\cal O}_{tG} +{\cal O}_{tG}^\dag \big) \quad \text{with}  \quad
 {\cal O}_{tG} =g_s \left(\bar Q_L \, \sigma^{\mu \nu} \, T^a\, t_R \right) \tilde{H}  G_{\mu\nu}^a \, ,
\ee
one finds 
\be
\mu = - \frac{\sqrt{2} m_t v}{\Lambda^2} c_{tG}\ ,
\ee
which implies that $c_{tG}/\Lambda^2 =0.1/[1 \text{TeV}]^2$ corresponds to $\mu =-0.006$. Notice the change of sign. Above,  $Q_L$ and $t_R$ stands for the $SU(2)_L$ left-handed doublet of top-bottom quarks and right-handed top quark fields respectively, while $\tilde{H}$ is as usual the dual of the $SU(2)_L$  doublet Higgs field, with SM vacuum expectation value $v$ given by $\langle 0| \tilde{H} |0\rangle =(v/\sqrt{2},0)$.

The addition of an effective magnetic dipole moment term to the SM Lagrangian, gives rise in general to further mixture
contributions, thus counteracting the generation of entanglement of the $t \bar{t}$ spin state produced by the SM interaction.
Specifically, using the same notations introduced in Section III.A and the coefficients collected in the Appendix,
in the $q \bar{q}$ production channel, again for transversally produced top quark pair ($\Theta=\pi/2$),
the two-spin density matrix can still be expressed as the convex
combination in (\ref{rho-qq}), but with the parameter $\lambda$ replaced by
\be
\tilde\lambda = \frac{\beta_t^2}{2-\beta_t^2 +9\mu f_{q\bar q}}\simeq \lambda - 
\mu \frac{9 f_{q\bar q}\beta_t^2}{(2-\beta_t^2)^2}\ ,
\quad \text{with} \quad f_{q\bar q}= \frac{N_c^2-1}{N_c^2}  \ .
\ee
At threshold, $\rho_{t\bar t}^{(q\bar q)}$ is still the totally mixed, unentangled state $\rho^{(1)}_{\rm mix}$ as before, but at 
high momentum, $\rho_{t\bar t}^{(q\bar q)}$ is no longer maximally entangled as $\tilde\lambda =1- 9\, \mu f_{q\bar q} <1$. 
Notice that the convexity condition $\tilde\lambda\in[0,1]$ requires $\mu\geq0$.

The case of the $gg$ production channel is  more involved. Nevertheless, at threshold the presence of the
magnetic dipole moment contribution is ineffective and $\rho_{t\bar t}^{(gg)}=\rho^{(-)}$, still maximally entangled.
On the other hand, for non-vanishing $\beta_t$, the spin density matrix
can be expressed, at least for $\beta_t\geq1/\sqrt{2}$, as the following mixture of four contributions:
\be
\rho_{t\bar t}^{(gg)}= \tilde{a}\rho^{(+)} + \tilde{b} \rho^{(-)} + \tilde{c} \rho^{(1)}_{\rm mix}+\tilde{d}\rho^{(3)}_{\rm mix}\ ,
\ee
where
\be
\tilde{a}=\frac{F_{gg}}{\tilde A}\beta_t^2\big(2\beta_t^2-1\big)\ ,\quad 
\tilde{b}=\frac{F_{gg}+7\mu f_{gg}}{\tilde A}(1-\beta_t^2)\ ,\quad 
\tilde{c}=\frac{4F_{gg}}{\tilde A}\beta_t^2(1-\beta_t^2)\ ,\quad 
\tilde{d}=\frac{7\mu f_{gg}}{\tilde A}\beta_t^2\ ,\quad 
\ee
with
\be
\tilde A=F_{gg}\big(1+2\beta_t^2 -2\beta_t^4\big) + 7\mu f_{gg}\ ,\quad \text{with} \quad F_{gg}=\frac{N_c^2-2}{64 N_c}\quad \text{and} \quad
\quad f_{gg}=\frac{1}{N_c (N_c^2-1)}\ . 
\ee

As $\beta_t\rightarrow 1$, $\rho_{t\bar t}^{(gg)}$ remains a mixture
of the density matrices $\rho^{(+)}$ and $\rho^{(3)}_{\rm mix}$, with mixing parameter 
$\tilde{a}=1-\tilde{d}=F_{gg}/(F_{gg} +7\mu f_{gg})$,
and thus with entanglement content no longer maximal. It is precisely the loss of entanglement induced by the presence
of a non-vanishing magnetic dipole moment contribution both in the $q \bar{q}$ and $gg$ production channels
that allows the  bound on the magnitude of the extra, effective parameter $\mu$, to be obtained.

As ${\cal L}_{\text{\tiny  dipole}}$ is here treated as a small perturbation to the SM Lagrangian, 
only the lowest order contributions in $\mu$ are retained in the evaluation of the top-pair spin state $\rho_{t\bar t}$ 
and in the entries of the corresponding correlation matrix $C_{ij}={\rm Tr}[\rho_{t \bar t}\, (\sigma_i\otimes\sigma_j)]$. As a consequence, the condition of positivity of the matrix $\rho_{t \bar t}$ might not be
automatically guaranteed, and therefore it needs to be imposed in order to get physically tenable results:
 this might lead to possible constraint on the range of the values that the parameter $\mu$ can take.
%This point has been overlooked in the discussions appeared in the recent literature.

%%%%%%%%%%%%%%
 \begin{figure}[h!]
\begin{center}
\includegraphics[width=3.3in]{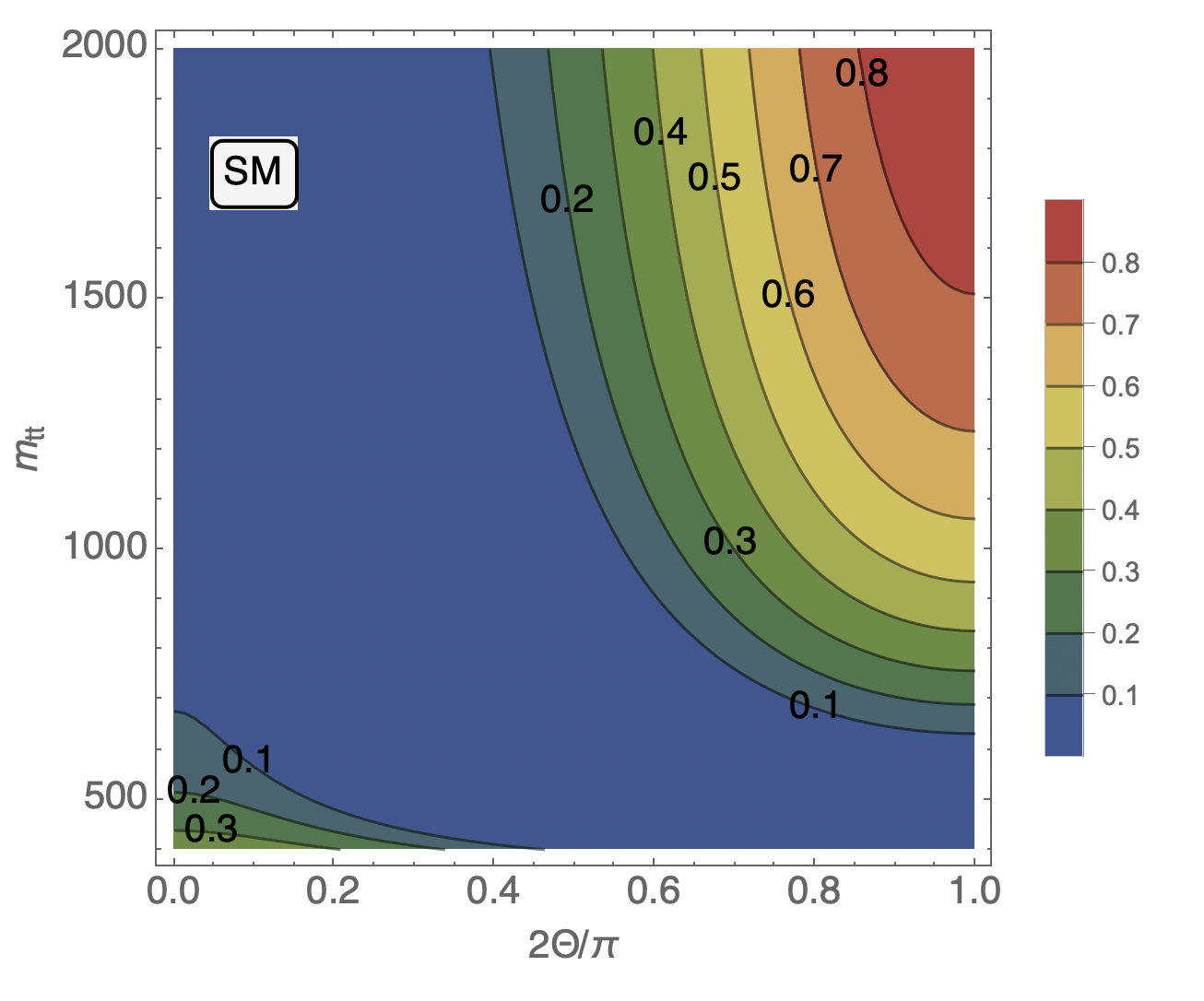}
\includegraphics[width=3.3in]{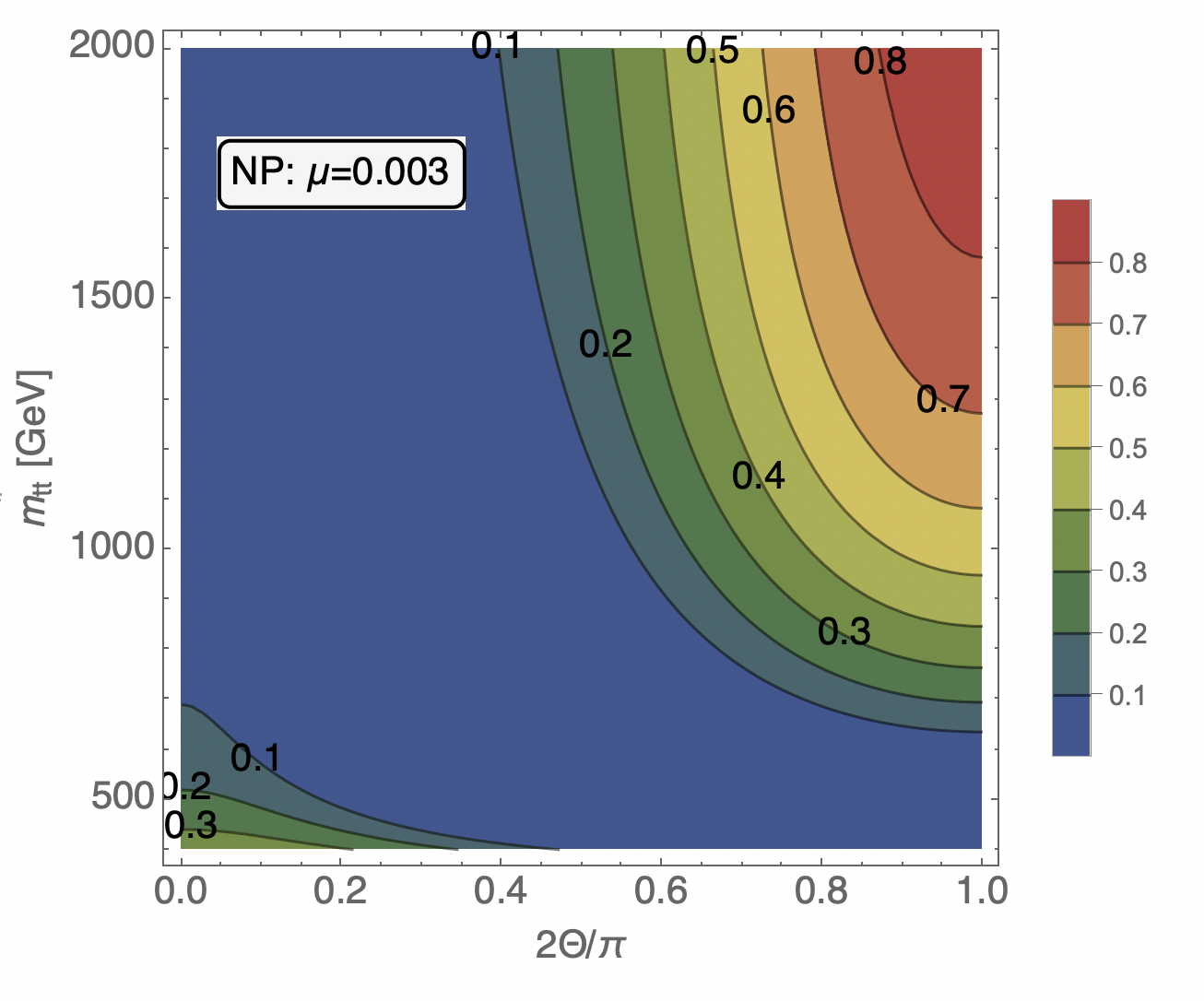}
\caption{\small \underline{On the left}: Concurrence ${\cal C}[\rho]$ in the SM as a function of the kinematical variables $\Theta$ and $m_{t \bar t}$. \underline{On the right}: Concurrence with new physics (NP): magnetic dipole moment with  $\mu=0.003$.
\label{fig:dipole} 
}
\end{center}
\end{figure}
%%%%%%%%%%%%%%%%%%%%%% 

The comparison between SM and new physics entanglement is best done by means of the concurrence observable ${\cal C}[\rho]$. As before, the presence of only one off-diagonal matrix element in the matrix $C$, as given in \eq{matrixC}, makes possible to write the concurrence in a simple analytic form as
\be
{\cal C}[\rho]=\frac{1}{2} \max \Big[ 0,\, | C_{rr} + C_{kk} | - (1-C_{nn}),\, \sqrt{(C_{rr}-C_{kk})^2 + 4 C_{rk}^2}- |1 - C_{nn}|\Big]
\ee

Fig.~\ref{fig:dipole} shows the concurrence in the whole kinematical region  for the SM and in the presence of the magnetic dipole operator for the  value of $\mu=0.003$. In both cases, the entanglement grows as we reach into larger energies and larger scattering angles.

\begin{figure}[h!]
\begin{center}
\includegraphics[width=1.9in]{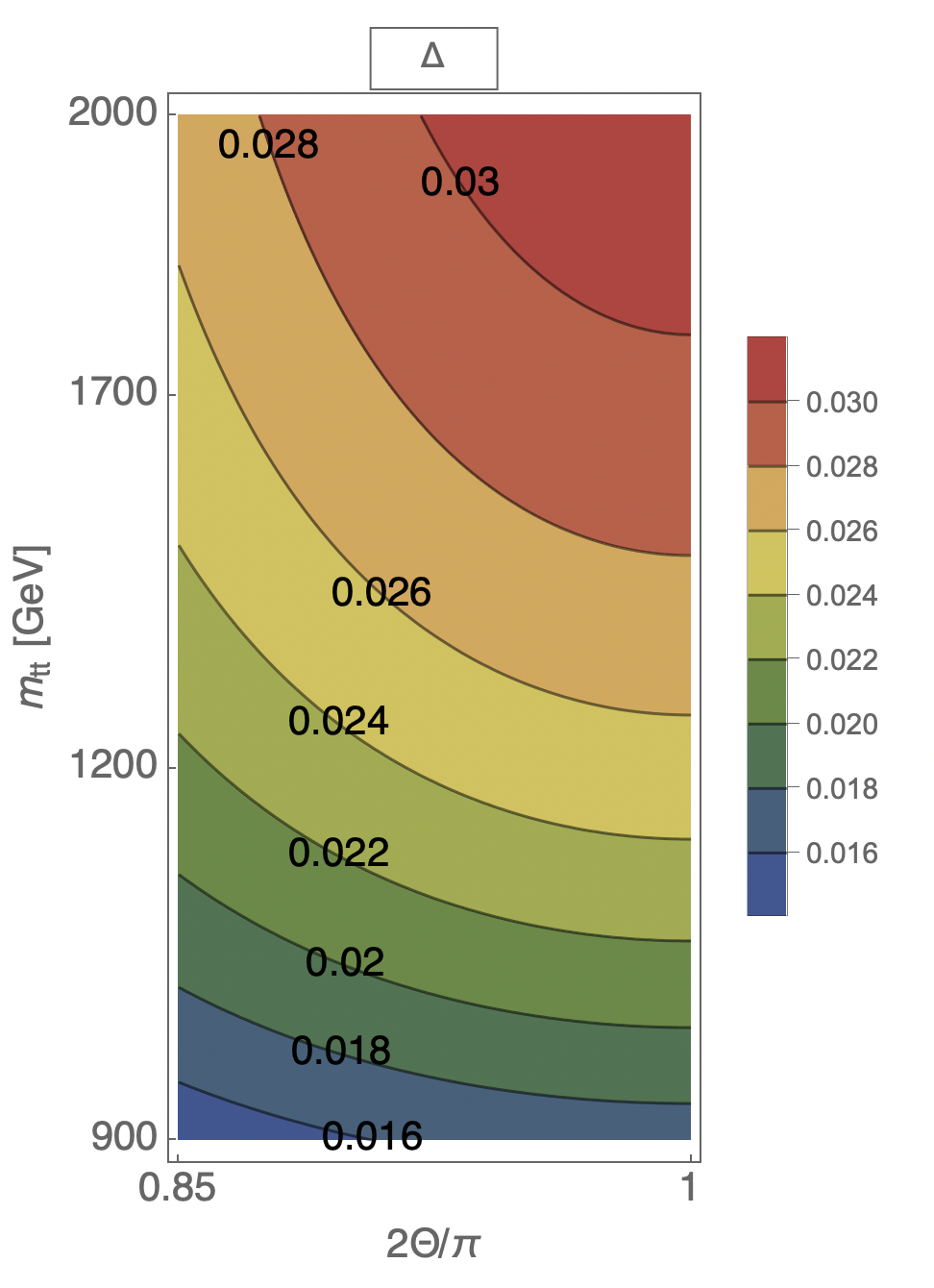}
\includegraphics[width=3.4in]{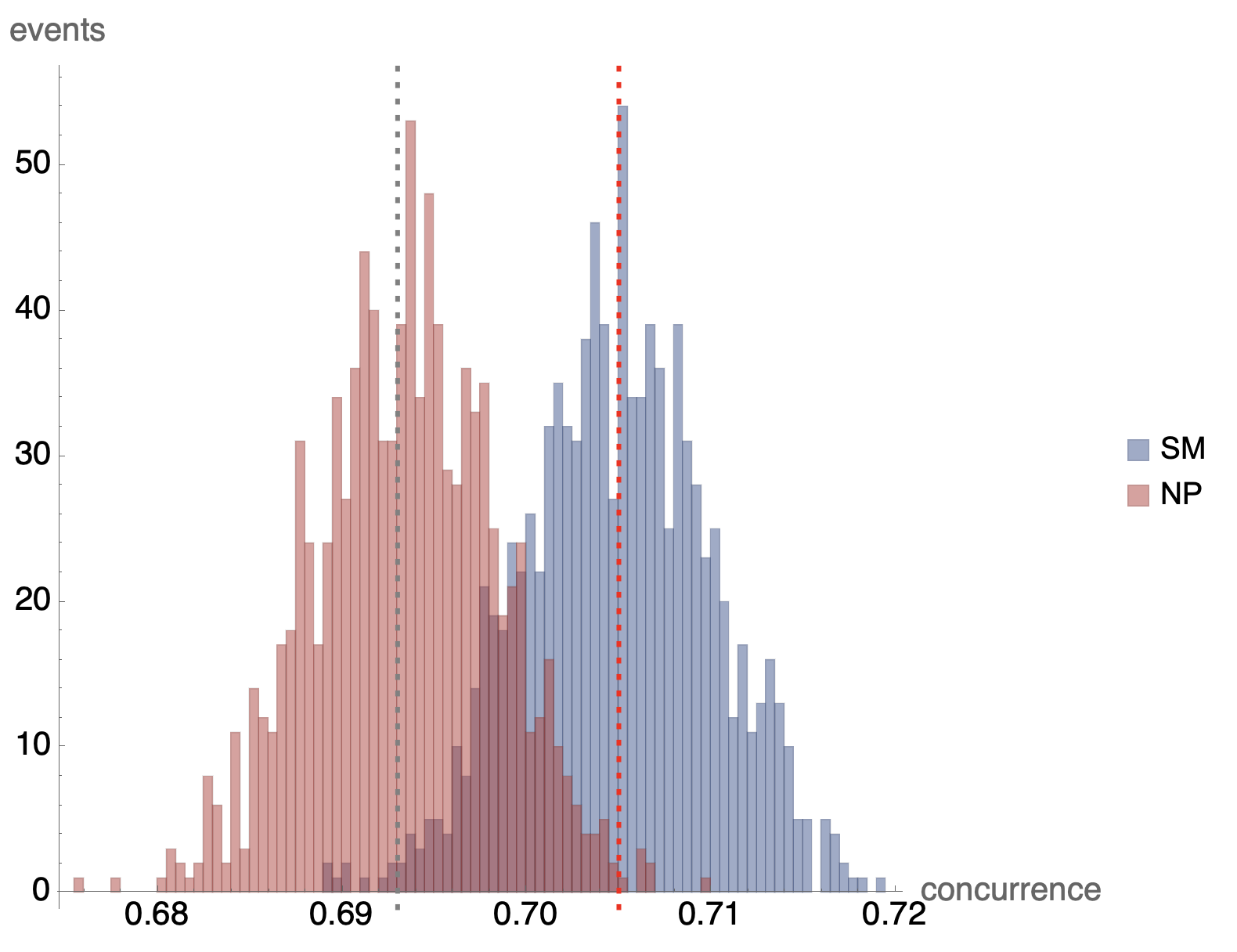}
\caption{\small \underline{On the left}: Percent difference in concurrence ${\cal C}[\rho]$ between SM and new physics ($\mu=0.003$) in the kinematical window $m_{t\bar t} > 900$ GeV and $2 \Theta/\pi > 0.85$ where $\Delta$ is defined as the difference between SM and new physics over the mean value of the SM. \underline{On the right}: The statistical distributions around the central values $0.705$ and $0.693$ for the SM and the new physics, respectively.
\label{fig:significanceTOP} 
}
\end{center}
\end{figure}

Fig.~\ref{fig:significanceTOP} shows the kinematical region $m_{t\bar t} > 900$ GeV and $2 \Theta/\pi > 0.85$ where the relative difference $\Delta$ between SM and new physics (with $\mu=0.003$) is largest and equal to about 3\%. This result is in agreement  with what found in~\cite{Aoude:2022imd} (with $c_{tG} =-0.1$ for $\Lambda=1 \,\text{TeV}$). We concentrate on this kinematical window to define the corresponding distribution of the values of the concurrence. The distribution for the SM has a central value $0.705$, that in  presence of the new physics becomes $0.693$.

As in the case of the operator $\mathfrak{m}_{12}[C]$, a segmentation of the window into smaller bins improves the separation between  SM and new physics and  can be implemented to further strengthen the bound. 

By running the toy Monte Carlo described in Section II for the distributions corresponding to SM and new physics, we find that, with the 463 events of LHC run2, a separation of 2.4$\sigma$ is  possible down to  the value of $\mu=0.003$.  

This result must be compared with current determinations~\cite{Lillie:2007hd,CMS:2019kzp} based on single observables which find  a weaker bound around  $\mu=0.02$. It is comparable to that obtained from the EFT global fit~\cite{Brivio:2019ius} which utilize multiple observables. We expect that when the new quantum probe of entanglement is added to the other data of the EFT global fit, it will  improve the overall sensitivity.

%%%%%%%%%%%%%%
%%%%%%%%%%%%%%%%%%%%%% 

\subsubsection{Consistency of the approximations}

The estimates we have performed are based on three approximations which must be verified for consistency. 

First of all, the linear approximation in the inclusion of the new physics is justified as long as the dipole operator, which scales with the energy of the process, is much smaller than the SM contribution. A rough estimate is provided by taking  $m_{t\bar t} \simeq 1000$ GeV  (the upper bound of our kinematical region) and $\mu\simeq 0.003$ (the benchmark value of the dipole). We have that the new physics term is order
\be
\frac{m_{t\bar t} \, \mu}{ 2 m_t} \simeq 0.01 \label{est}
\ee
smaller than the SM and therefore the linear approximation of retaining only single insertions of the dipole operator seems to be justified. 

 Since the chromo-magnetic operator is an effective operators of dimension 5, its effect grows with the energy.  The relevance of  inserting twice this operator with respect to the single insertion becomes relevant only in kinematic regions of top-pair invariant masses that are  close to  the breaking of the perturbative expansion of the effective theory. We have   restricted our analysis to effective scales below such a region.   In particular, in the plot of Fig. 5, the limit  $\mu=0.003$ corresponds to an effective scale associated to the
chromo-magnetic operator  of order 100 TeV, which is much above the maximum range considered $m_{t\bar{t}} \sim 2 \;{\rm TeV}$. 
Then, for invariant masses $m_{t\bar{t}} <  O (1 \;{\rm TeV})$, the quadratic contributions of the magnetic-dipole operator is expected to be negligible because proportional to terms of the order of $O (m^2_{t\bar{t}}/\Lambda^2)$. Accordingly, quadratic corrections are expected to be very small.  This is confirmed by the analysis in~\cite{Aoude:2022imd} at the LO which  shows that the effect of terms quadratic  in the dipole operator is negligible in the kinematical region we have considered. A NLO computation appeared recently in \cite{Severi:2022qjy}.

Secondly,  the same estimate in \eq{est} also shows that the  SMEFT operator expansion is justified even with an operator like the magnetic dipole that grows with the CM energy. At least for CM energies up to and around 1 TeV the correction is perturbative.

Finally, a source of concern comes about the size of QCD NLO terms with respect to the new physics term.
The one-loop QCD corrections give rise to a dipole operator with coefficient
\be
- \frac{\alpha_s}{\pi} \frac{m_t^2}{m_{t \bar t}^2} \log \frac{m_{t \bar t}^2}{m_t^2} 
\ee
which, in the relevant kinematical window, is  about 4 times smaller than the NP term with $\mu=0.003$. Therefore, the QCD contribution to the dipole operator can be neglected, as we did, but  a full QCD NLO estimate, though computationally challenging, will be necessary if  the limit is to be strengthened.

\subsection{The semi-leptonic decays of the top quark}

The above analysis is done on the pseudo-observables defined in terms of the top-quark pairs. In the final analysis, these must be computed in terms of actual observables, namely the momenta of the final leptons coming from the decay of the top quarks.
Introducing  the  angles
\be
\cos \theta^a_+ = \hat{\ell}_+\cdot \hat{a}\quad \text{and} \quad \cos \theta^b_- = \hat{\ell}_-\cdot \hat{b}\, ,
\ee
where  $\hat{b}=-\hat{a}$ and $\hat{a} \in \lbrace \hat{k},~ \hat{r}, ~\hat{n} \rbrace$,
the cross section is given by
\be
\frac{1}{\sigma} \frac{\di \sigma}{\di \Omega_+ \di \Omega_-} = \frac{1}{16 \pi^2} \Big( 1 + B_- \cdot \hat{\ell}_+ + B_+ \cdot \hat{\ell}_- -  \hat{\ell}_- \cdot C \cdot \hat{\ell}_+ \Big)
\ee
so that, in the absence of acceptance cuts, the elements of the matrix $C$ can be expressed \cite{Bernreuther:2015yna} as 
\be
C_{ab} \big[ m_{t\bar{t}}, \cos \Theta \big] = - 9 ~\dfrac{1}{\sigma} \int d\xi_{ab}  \dfrac{d \sigma}{d \xi_{ab}} \xi_{ab}\, ,
\label{eq:Cexp}
\ee
where we defined
\be
\xi_{ab} = \cos \theta^a_+ \cos \theta^b_- \, ,
\ee
and with the residual dependence of the cross section $\sigma$ on intrinsic kinematic variables made explicit in the argument of the matrix elements.

Such analysis based on the actual data can only be done by the experimental collaborations. It requires the full simulation of the detector, an estimate of systematic errors and the reconstruction efficiency. We are aware that a significant deterioration in significance will take place.

\section{Results: tau lepton pairs} 

The case in which the states produced by the interaction are $\tau$ leptons can be discussed along the same lines as for the top quarks. The dominant process is the Drell-Yan production in which the quarks go  either into a photon or a $Z$-boson which, in turn, decay into the $\tau$-lepton pair. In addition to the production, in this case we also have the  process in which the $\tau$ leptons originate from the Higgs boson decay. We discuss the possible role of quantum entanglement in both these two physics processes.

\subsection{Drell-Yan}

The production of $\tau$-lepton pairs via Drell-Yan in the SM receives contributions from the $s$-channel photon, the $Z$-boson and their interference. They provide an ideal laboratory for studying entanglement. Because the fewer the contributions, the larger the entanglement  (as mixing diminishes quantum correlations), we expect this to be larger at low-energies (where the photon diagram dominates) or around the $Z$-boson pole (where the $Z$-boson diagram dominates).  
At low energies, the cross section is dominated by the photon term which produces entangled $\tau$-lepton pairs, while at high-energies all terms contribute and entanglement is suppressed. Around the $Z$-boson pole the cross section is dominated by the corresponding term with maximal entanglement.

\subsection{Entanglement in $\tau\bar \tau$ production}

As in the case of the top-quark pair production, the two spin-1/2 state is described by
a density matrix of the general form (\ref{rho}),
whose entries depend on the kinematic variable $\beta_\tau=\sqrt{1-4m_\tau^2/m_{\tau\bar\tau}^2}$, 
with $m_{\tau\bar\tau}$ the $\tau$-pair invariant mass,
and on the scattering angle $\Theta$. 

Using the same reference frame and notation introduced as for the top pair production in Section III.A,
and again focusing on the situation
of transversally produced lepton pairs ($\Theta=\pi/2$), as previously mentioned, one can distinguish three kinematical regions: The first,
at low energies, $m_{\tau\bar\tau}\ll m_Z$, where photon exchange is dominating, the intermediate one, 
$m_{\tau\bar\tau}\simeq m_Z$, dominated by the $Z$ exchange and finally the high energy one, $m_{\tau\bar\tau} \gg m_Z$. 

With the help of the results collected in the Appendix, in the low-energy regime (for which $m_{\tau\bar\tau}\ll m_Z$) and for all quark production channels,
the $\tau$-pair spin state can be represented by the convex combination as in (\ref{rho-qq}),  
\be
\rho_{\tau\bar\tau}= \lambda \rho^{(+)} + (1-\lambda)\rho^{(1)}_{\rm mix} \quad \text{with} \quad  \lambda=\frac{\beta_\tau^2}{2-\beta_\tau^2}\in [0,1]\, ;
\ee
at threshold, $\beta_\tau\simeq 0$, the state is a totally mixed one, with no quantum correlations, while as 
$\beta_\tau \rightarrow 1$, the spins of the $\tau$-lepton pair tend to be generated in a maximally entangled state.

As the $Z$-channel starts to become relevant, this entanglement is however progressively lost due to the mixing
between the photon and $Z$ contribution. Nevertheless, a revival of entanglement occurs as the $Z$ channel become dominant,
$m_{\tau\bar\tau}\simeq m_Z$; in this region, with the notation and conventions introduced in the Appendix,
the two-spin density matrix can be described by the following convex combination,
for all quark production channels :
\be
\rho_{\tau\bar\tau}= \lambda \tilde\rho^{(+)} + (1-\lambda)\tilde\rho^{(2)}_{\rm mix}\ ,\qquad 
\lambda=\frac{(g_A^\tau)^2 - (g_V^\tau)^2 }{(g_A^\tau)^2 + (g_V^\tau)^2 }\ ,
\ee
where, 
\be
\tilde\rho^{(2)}_{\rm mix}=\frac{1}{2}\Big( |\text{\small RR}\rangle\langle \text{\small RR}| 
+ |\text{\small LL}\rangle\langle \text{\small LL}| \Big)\ ,\\
\ee
while,
\be
\tilde\rho^{(+)}= |\tilde\psi^{(+)}\rangle\langle\tilde\psi^{(+)}|\ ,\qquad 
|\tilde\psi^{(+)}\rangle=\frac{1}{\sqrt{2}}\Big( |\small{+-}\rangle + |\small{-+}\rangle \Big)\ ,
\ee
is a projector on a Bell state as in (\ref{Bell-states}), but now expressed in terms of the eigenvectors of $\sigma_x$. 
It turns out that $\lambda\simeq 1$, so that $\rho_{\tau\bar\tau}$ is very close to
the maximally entangled state $\tilde\rho^{(+)}$.

Finally, in the high energy regime ($m_{\tau\bar\tau}\gg m_Z$)  both photon and $Z$ channel contribute, and their mixing
lead to a rapid depletion of entanglement. Indeed, for each $q\bar q$ production channel, the $\tau$-pair spin correlations
can be described in terms of the following density matrix:
\be
\rho_{\tau\bar\tau}= \lambda^q \rho^{(+)} + (1-\lambda^q)\tilde\rho^{(2)}_{\rm mix}\ ,
\qquad \lambda^q=\frac{1-R_-^q}{1+R_+^q}\ ,
\label{rho-high}
\ee
where $\rho^{(+)}$ is as in (\ref{Bell-states}), while
\be
R^q_\pm = \frac{\chi^2(m_{\tau\bar\tau}^2) \big[ (g_A^q) + (g_V^q) \big] \big[(g_A^\tau) \pm (g_V^\tau)\big]}
{(Q^q)^2 (Q^\tau)^2 + 2\,{\rm Re}\chi(m_{\tau\bar\tau}^2)\, Q^q Q^\tau\,  g_V^q g_V^\tau}\ .
\ee
Specifically, in the case of the $u$ quark production channel, one finds $\lambda^u\simeq 0.7$, 
so that some entanglement is preserved,
while for the $d$ quark production channel, as $\lambda^d\simeq 0.1$, entanglement is essentially lost.

For completeness, it should be noticed that each $\tau$ lepton
is produced in a partially polarized state, as some of the single-spin polarization coefficient $B_i^\pm$ 
in (\ref{rho}) are non-vanishing (see Appendix).
This is particularly relevant for the quark $d$ production channel, where the magnitude
of these single particle terms is of the same order of the entries of the correlation matrix $C_{ij}$,
while for the $u$ production channel they are about one order of magnitude smaller.
This implies that the full density matrix describing the $\tau$-pair spin state $\rho_{\tau\bar\tau}$ is really in this case
a mixture of (\ref{rho-high}) with additional states
further reducing in general its entanglement content.

In addition, as discussed below, the full correlation matrix $C_{ij}$ is obtained by putting together all
relevant $q\bar q$-production channel contributions, weighted by suitable luminosity functions and with appropriate
normalization: this leads to further mixing and in general to additional loss of entanglement.

\subsection{ Computing the observables}

As before,   we compute all the entries of the correlation matrix $C_{ij}$ from the process
\be
p+p \to \tau^- + \tau^+  \, .
\label{eq:process}
\ee
with the unpolarized cross section given by
\be
\frac{\di \sigma}{\di \Omega \, \di m_{\tau\bar\tau}} = \frac{\alpha^2 \beta_\tau}{64 \pi^2 m_{\tau \bar \tau}^2} \Big\{ L^{uu} (\tau) \, \tilde A^{uu}[m_{\tau\bar\tau},\, \Theta]+\big[L^{dd} (\tau)+L^{ss}(\tau) \big]\, \tilde A^{dd}[m_{\tau\bar\tau},\, \Theta]  \Big\}
\ee
with  $L^{qq}(\tau)$ the parton luminosity functions of section II,  $\tau = m_{\tau^- \tau^+}/\sqrt{s}$  and $\alpha=e^2/4 \pi$. The explicit expressions for $\tilde A^{uu,dd}(m_{\tau\bar\tau})$  are given in the Appendix. 

%%%%%%%%%%%%%%%%%%%%%%%%%%%%%%%%%%
 \begin{figure}[h!]
\begin{center}
\includegraphics[width=5in]{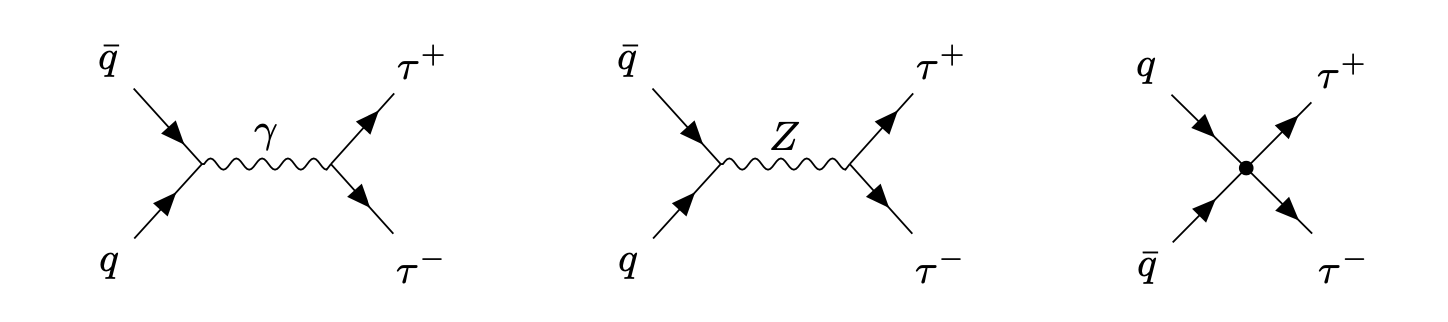}
\caption{\small \label{fig:taus_cont} Feynman diagrams for $\tau^-\tau^+$ production. \underline{On the right}, the contact interaction (see \eq{CI}).}
\end{center}
\end{figure}
%%%%%%%%%%%%%%%%%%%%%%%%%%%%%%%%%%%%%%%%%%%%%

%%%%%%%%%%%%%%
 \begin{figure}[h!]
\begin{center}
\includegraphics[width=3in]{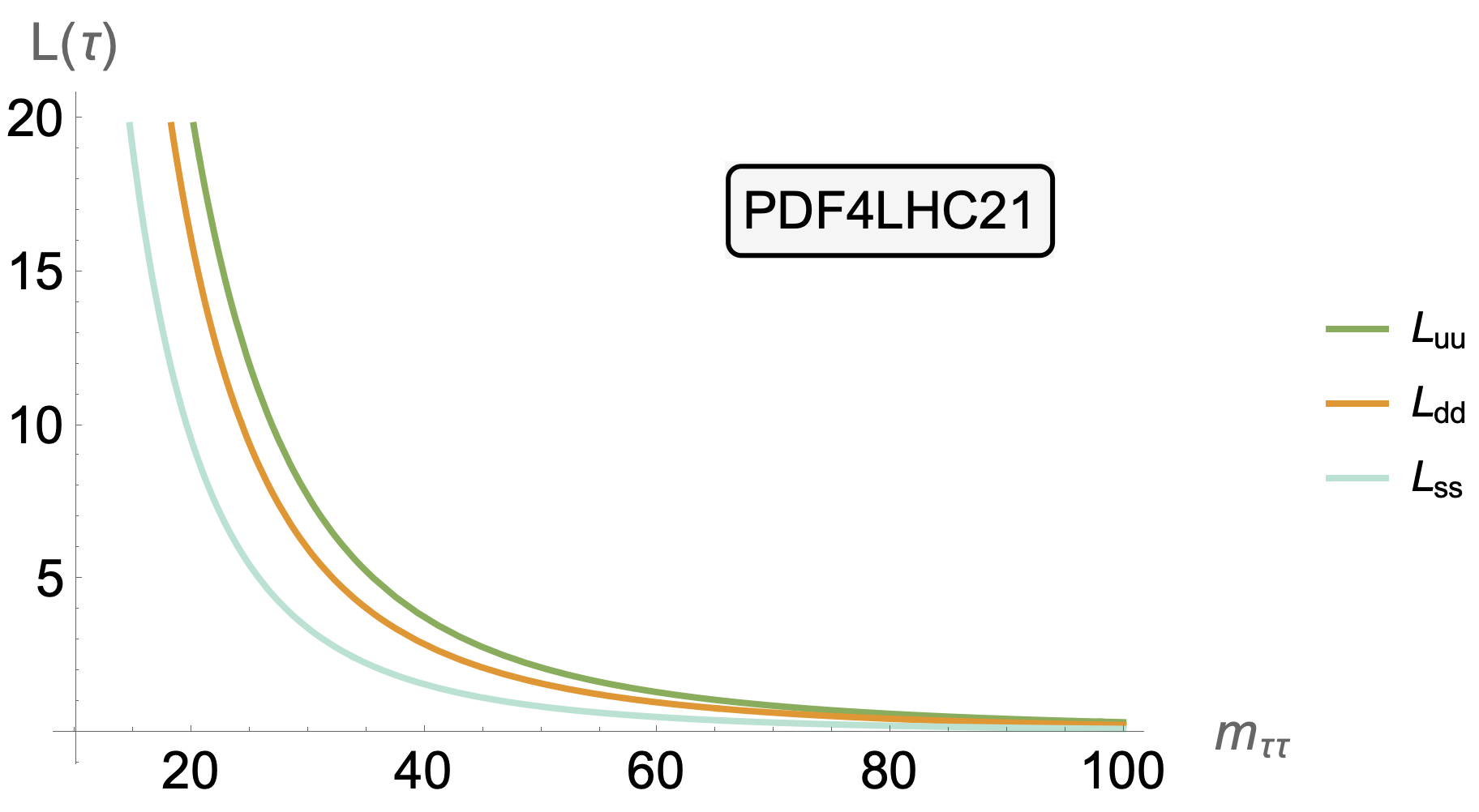}
\includegraphics[width=3.1in]{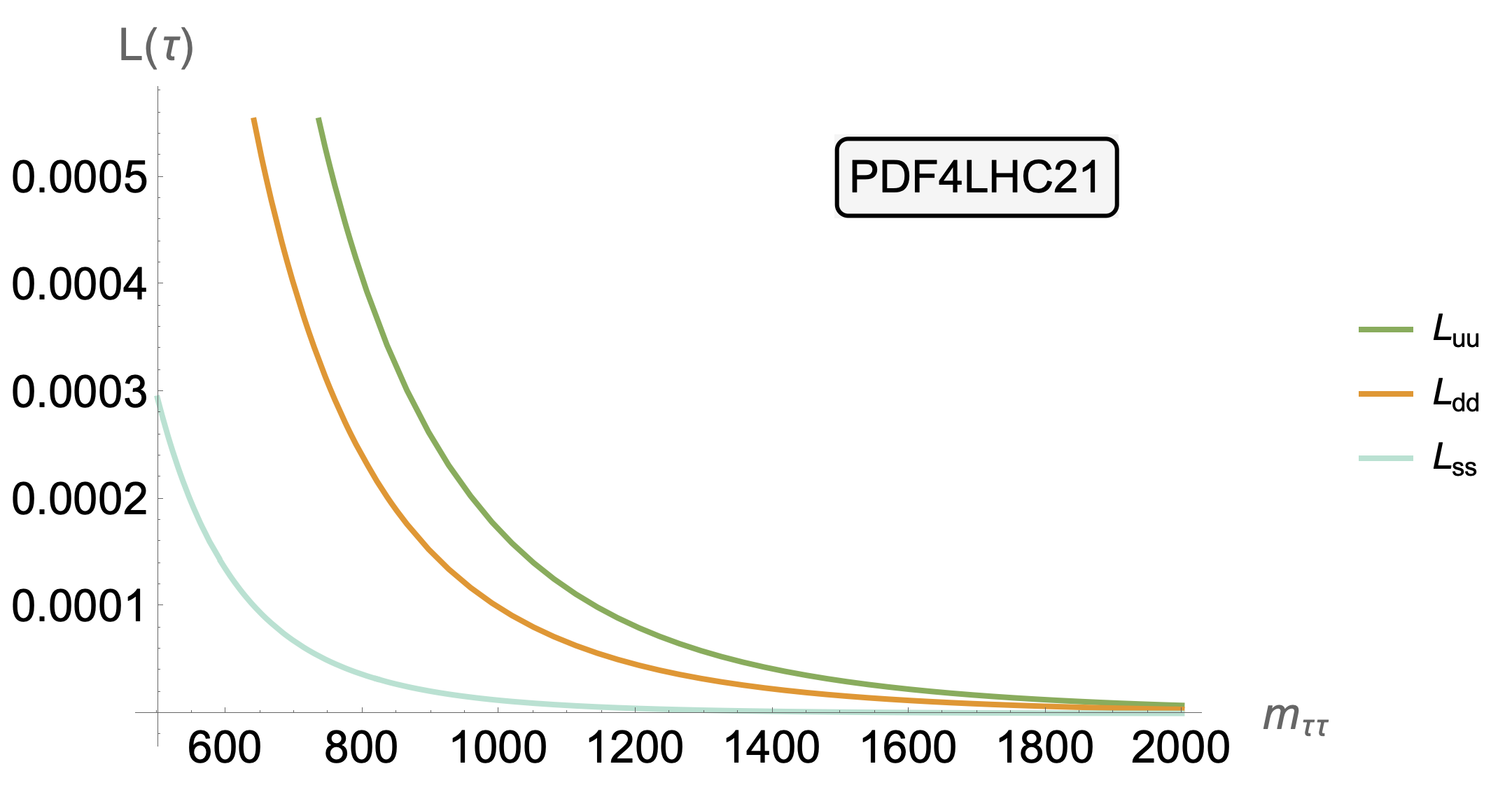}
\caption{\small  Parton luminosity functions: The up quark luminosity is  about 28\% larger than the down quark luminosity at threshold, then increasing  to about 38\%  around the $Z$-boson pole and eventually reaching 70\% around 1 TeV.
\label{fig:pdftau} 
}
\end{center}
\end{figure}
%%%%%%%%%%%%%%%%%%%%%% 

We combine the two channels (see Fig.~\ref{fig:taus_cont}) $u+\bar u\to \tau^- + \tau^+$ (which enters with  factors function of  $Q_u=2/3$) and $d +\bar d \to \tau^- \tau^+$ and $s +\bar s \to \tau^- \tau^+$ (which enters with  factors function of  $Q_d=-1/3$) by weighting the respective contributions through the parton luminosity functions
\be
L^{qq} (\tau)= \frac{4 \tau}{\sqrt{s}} \int_\tau^{1/\tau} \frac{\di z}{z} q_{q} (\tau z) q_{\bar q} \left( \frac{\tau}{z}\right)
\ee
where the $q(x)$ are the PDFs for respectively the $u$, $d$ and $s$ quarks.  As before, their numerical values are those provided by   {\sc PDF4LHC21}~\cite{PDF4LHCWorkingGroup:2022cjn} for $\sqrt{s}=13$ TeV and factorization scale $q_0=m_{\tau \bar \tau}$ (see Fig.~\ref{fig:pdftau}).

Therefore, we have that
\be
C_{ij} [m_{t\bar t},\, \Theta]= \frac{L^{uu} (\tau) \, \tilde C_{ij}^{uu}[m_{\tau\bar\tau},\, \Theta]+\big[L^{dd} (\tau)+L^{ss} (\tau) \big]\, \tilde C_{ij}^{dd}[m_{\tau\bar\tau},\, \Theta]} {L^{uu}(\tau) \, \tilde A^{uu}[m_{\tau\bar\tau},\, \Theta]+\big[L^{dd} (\tau)+L^{ss} (\tau)\big] \, \tilde A^{dd}[m_{\tau\bar\tau},\, \Theta]} \label{pdf2}\, ,
\ee
where the down-quark luminosities can be grouped together because they multiply the same correlation functions.
The expression in \eq{pdf2} must be expanded in the case of new physics by retaining the linear terms.

\subsection{Bell inequalities}

\begin{figure}[h!]
\begin{center}
\includegraphics[width=3.3in]{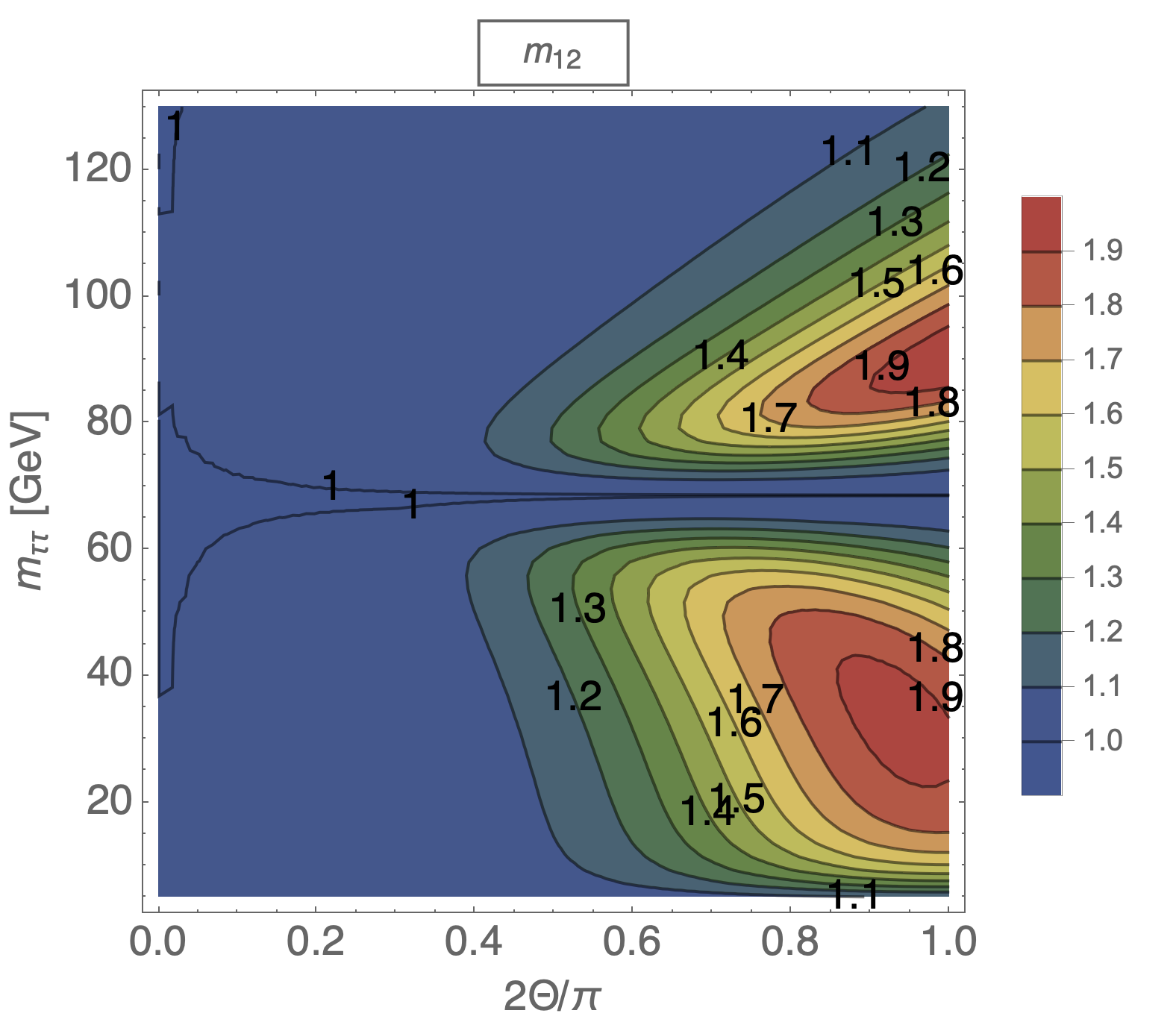}
\caption{\small  Eigenvalues $\mathfrak{m}_{12}[C]$ as a function of the kinematical variables $\Theta$ and $m_{t \bar t}$ across the entire available space. 
\label{fig:m1m2Atau} 
}
\end{center}
\end{figure}

The values of the observable $ \mathfrak{m}_{12}[C]$ are shown in Fig.~\ref{fig:m1m2Atau} across the entire kinematical space.  The figure confirms the qualitative analysis of section IV.B. It shows that, for large scattering angles, entanglement is close to maximal (that is,$ \mathfrak{m}_{12}[C]$ close to 2) 
where the invariant mass of the $\tau$-lepton pairs selects one of the two possible channels with either the photon or the $Z$-boson exchange dominating.

%%%%%%%%%%%%%%
%%%%%%%%%%%%%%%%%%%%%% 

%%%%%%%%%%%%
\begin{table}[h!]
\bc
\begin{tabular}{ccc}
%\toprule
&\hskip0.5cm (run 2) ${\color{oucrimsonred} {\cal L}=140\ \text{fb}^{-1}}$  \hskip0.5cm &  \hskip0.5cm (Hi-Lumi) ${\color{oucrimsonred} {\cal L}= 3\ \text{ab}^{-1}}$  \hskip0.5cm \\[0.2cm]
\hline\\
 \underline{events} \hskip0.4cm  &\hskip0.4cm $1.1 \times 10^6$  \hskip0.4cm &\hskip0.4cm $2.2 \times 10^7$ \hskip0.4cm \\[0.4cm]
   \hline%
\end{tabular}
\caption{\label{tab:events_taus_bell}Number of expected events in the kinematical region $20<m_{\tau \bar \tau} < 45$ GeV and $2 \Theta/\pi > 0.80$ }
\ec
\end{table}
%%%%%%%%%%%%%%%%%%

\begin{figure}[h!]
\begin{center}
\includegraphics[width=1.9in]{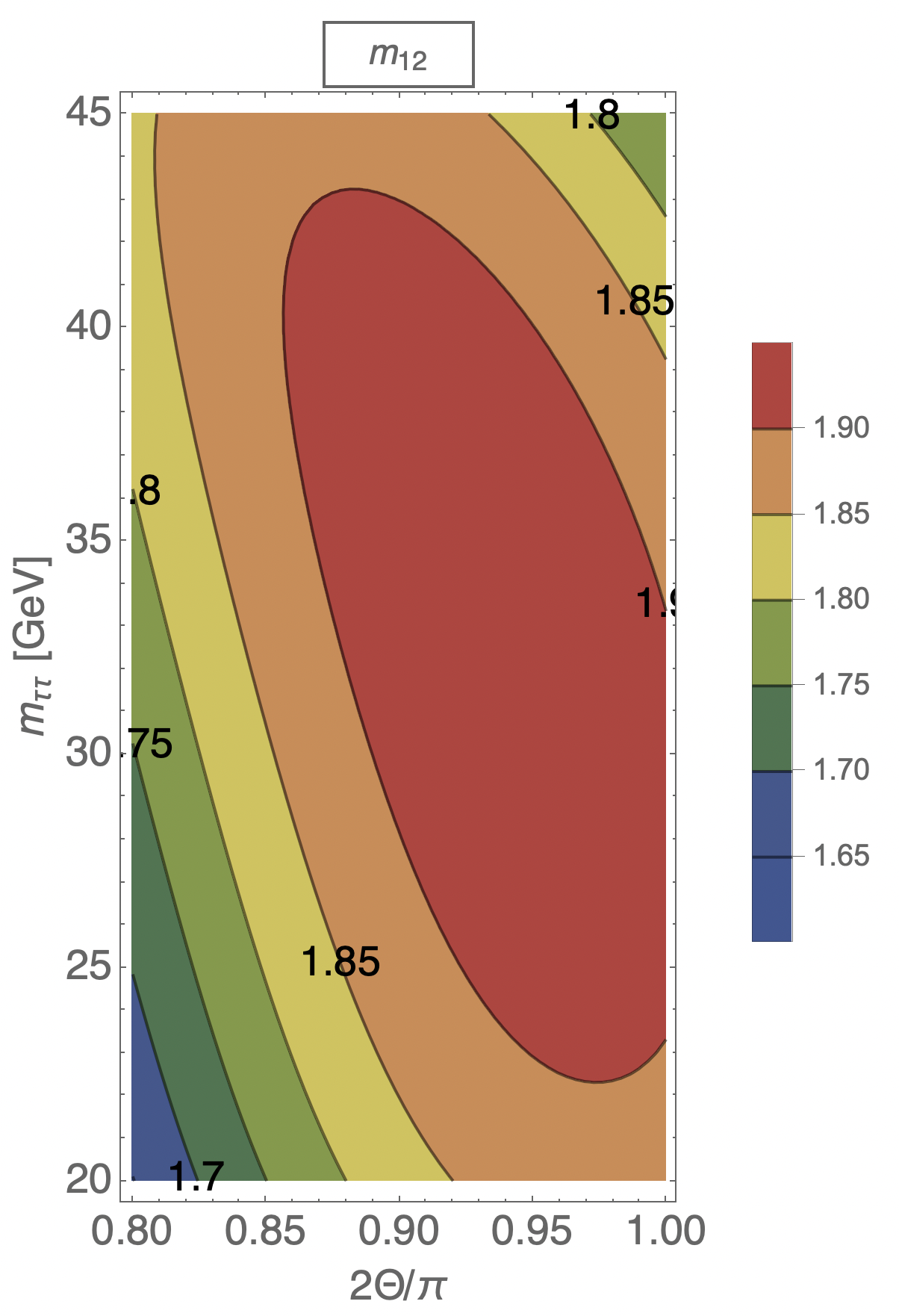}
\includegraphics[width=2.8in]{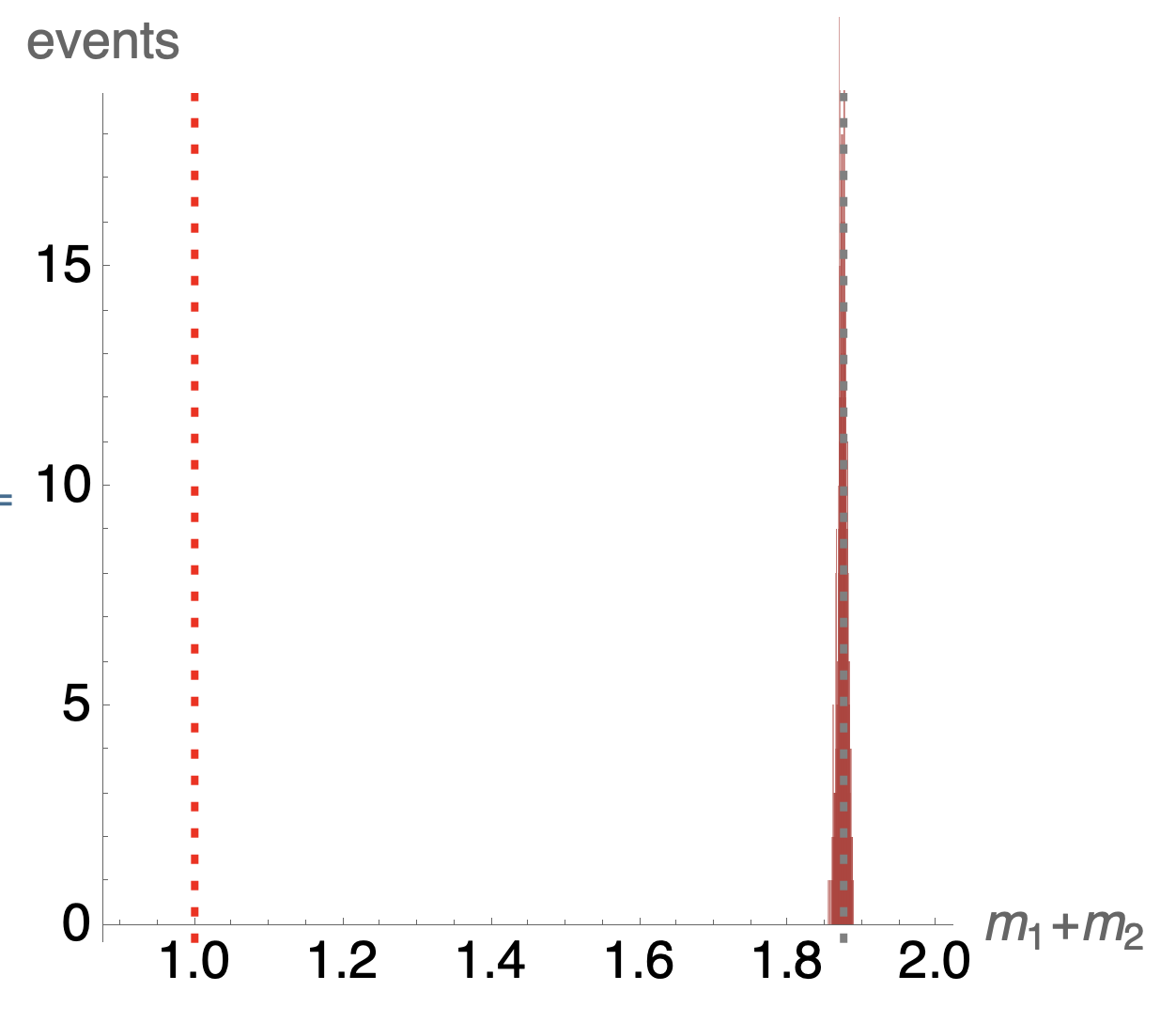}
\caption{\small \underline{On the left}: The kinematical window ($20<m_{\tau \bar \tau} < 45$, $2 \Theta/\pi > 0.80$) where the Bell inequality is to be tested.  \underline{On the right}: The statistical distribution for 100 events around the mean value $1.88$ with dispersion given by $\sigma = 0.006$.
\label{fig:significanceBellTau} 
}
\end{center}
\end{figure}

We take the kinematical window where $20<m_{\tau \bar \tau} < 45$ and $2 \Theta/\pi > 0.80$ as the most favorable to test the Bell inequalities and there estimate the operator $\mathfrak{m}_{12}[C]$. In this window the mean value of $\mathfrak{m}_{12}[C]$ is 1.88. The number of expected events at the LHC is large (see Table~\ref{tab:events_taus_bell}  in which the cross sections are computed by running {\sc MADGRAPH5}~\cite{Alwall:2014hca} at the LO and then correcting by the $\kappa$-factor given at  the NNLO~\cite{Grazzini:2017mhc}) and we  show the statistical significance of the hypothesis $\mathfrak{m}_{12}[C]> 1$ (as obtained by running the toy Monte Carlo described in section II)  for the case of having just 100 events (Fig.~\ref{fig:significanceBellTau}). The statistical significance is huge: more than 100. Indeed, this seems to be the process where the experimental confirmation of the violation of Bell inequalities is most likely thanks to the large number of events available (as opposed to the scarcity of events with the required energy in the case of the top quark).

  The main source of  theoretical uncertainty on the entanglement observables of the $\tau$ lepton pairs comes from the choice of the PDF, which is negligible in the relevant kinematic regions and giving an effect of the order of a per mille. As it was for the case of the top-quark pairs, here too the possible backgrounds are negligible once  the tagging on the exclusive decays modes of the $\tau$  and the kinematic cuts to reconstruct its on-shell mass are employed.

%%%%%%%%%%%%%%

\subsection{New physics: Contact interactions}

 The simplest new physics that can enter the Drell-Yan process is a contact interaction among quarks and $\tau$-leptons. Such a contact interaction mediates a process in which the quarks directly produce the leptons (see Fig~\ref{fig:taus_cont}). 
 
 The most general contact operators for the production of $\tau$-leptons from quarks can be written, in chiral components, as
 \begin{multline}
 {\cal L}_{\text{\tiny  cc}} = -\frac{4\pi }{\Lambda^2} \eta_{LL}  ( \bar q_L \gamma^\alpha q_L) \,(\bar \tau_L \gamma_\alpha \tau_L) \\
 -\frac{4\pi}{\Lambda^2} \eta_{RR} (\bar q_R \gamma^\alpha q_R)\, (\bar \tau_R \gamma_\alpha \tau_R)
 -\frac{4\pi}{\Lambda^2}  \eta_{LR} ( \bar q_L \gamma^\alpha q_L) \,(\bar \tau_R \gamma_\alpha \tau_R )
 -\frac{4\pi}{\Lambda^2} \eta_{RL} (\bar q_R \gamma^\alpha q_R)\, (\bar \tau_L \gamma_\alpha \tau_L) \label{CI} \, .
 \end{multline}

  These contact operators can be thought, in a UV complete theory, as  arising from the exchange of a scalar lepto-quark. The scale $\Lambda$ in \eq{CI} could be of the same order of the mass scale of the corresponding new physics particles. The exchange of a vector lepto-quark leads to  contact operators with no interference terms with the SM diagrams, in the quark and lepton massless limit, and therefore less constrained. The scale $\Lambda$ controls the size of these new-physics terms; the factor $4\pi$ is conventional and is there to remind us that the UV physics could come from a strong-coupling regime.

To gauge the effect of these terms is sufficient to take one of them. The function $\tilde A$ and $\tilde C_{ij}$ for the operator $LR$ (the third term in \eq{CI}) are given in Appendix with $\eta_{LR} = 1$. 

 Based on general grounds, all  operators in \eq{CI} give the same order of magnitude effect. Indeed, in model independent analysis, it is customary to assume the contributions from different operators to be uncorrelated, due to different potential NP contributions to each of the corresponding Wilson coefficients. Therefore, the bounds obtained from each separate contribution to the entanglement, are expected to be  the same order of the other operators, which only differs for a different chirality structure. 

The addition of such an effective contact interaction term to the SM Lagrangian modifies the picture we drew about the entanglement of the $\tau$-lepton pairs in Section IV.B
For a given quark $q\bar q$ production channel and in the high energy regime ($m_{\tau\bar\tau}\gg m_Z$),
the $\tau$-pair spin correlations can again be described in terms of the convex
combination in (\ref{rho-high}), but with the parameter $\lambda^q$ replaced by 
\be
\tilde\lambda^q = \lambda^q \Bigg[1+\eta\,\Bigg(\frac{\tilde R^q_-}{1+R_+^q} - \frac{\tilde R^q_+}{1-R_-^q}\Bigg)\Bigg]\ ,\quad  
\tilde R^q_\pm= \frac{Q^q Q^\tau\pm\chi(m_{\tau\bar\tau}^2) \big[ (g_A^q) + (g_V^q) \big] \big[(g_A^\tau) \pm (g_V^\tau)\big]}
{(Q^q)^2 (Q^\tau)^2 + 2\,{\rm Re}\chi(m_{\tau\bar\tau}^2)\, Q^q Q^\tau\,  g_V^q g_V^\tau}\ .
\ee
where  $\eta=m_{\tau\bar\tau}^2/\alpha\Lambda^2$.
For the $u$-quark production channel $\tilde\lambda^u \simeq \lambda^u \big(1+\eta/4\big)$,
while for the $d$-quark production channel one finds $\tilde\lambda^d \simeq \lambda^d \big(1-3\eta\big)$.
As both $\lambda^u$ and $\lambda^d$ are less than 1 and $\eta$ is small, one can get both an increase
and a decrease of the $\tau$-pair spin correlations according to the sign of $\eta$,
without violating the requirement of the positivity of the density matrix $\rho_{\tau\bar\tau}$.
It is precisely this change in the entanglement content of the $\tau$-pair spin state induced by the presence
of the contact term contribution, both in the $u \bar{u}$ and $d \bar{d}$ production channels,
that makes possible  obtaining bounds on the magnitude of the new physics scale $\Lambda$.

%%%%%%%%%%%%
\begin{table}[h!]
\bc
\begin{tabular}{ccc}
%\toprule
&\hskip0.5cm (run 2) ${\color{oucrimsonred} {\cal L}=140\ \text{fb}^{-1}}$  \hskip0.5cm &  \hskip0.5cm (Hi-Lumi) ${\color{oucrimsonred} {\cal L}=3\ \text{ab}^{-1}}$  \hskip0.5cm \\[0.2cm]
\hline\\
 \underline{events} \hskip0.4cm  &\hskip0.4cm  $27$  \hskip0.4cm &\hskip0.4cm  $573$ \hskip0.4cm \\[0.4cm]
   \hline%
\end{tabular}
\caption{\label{tab:events_taus}Number of expected events in the kinematical region $m_{\tau^+\tau^-}>800$ GeV and $0.85<x<1$.}
\ec
\end{table}
%%%%%%%%%%%%%%%%%%

%%%%%%%%%%%%%%
 \begin{figure}[h!]
\begin{center}
\includegraphics[width=3.3in]{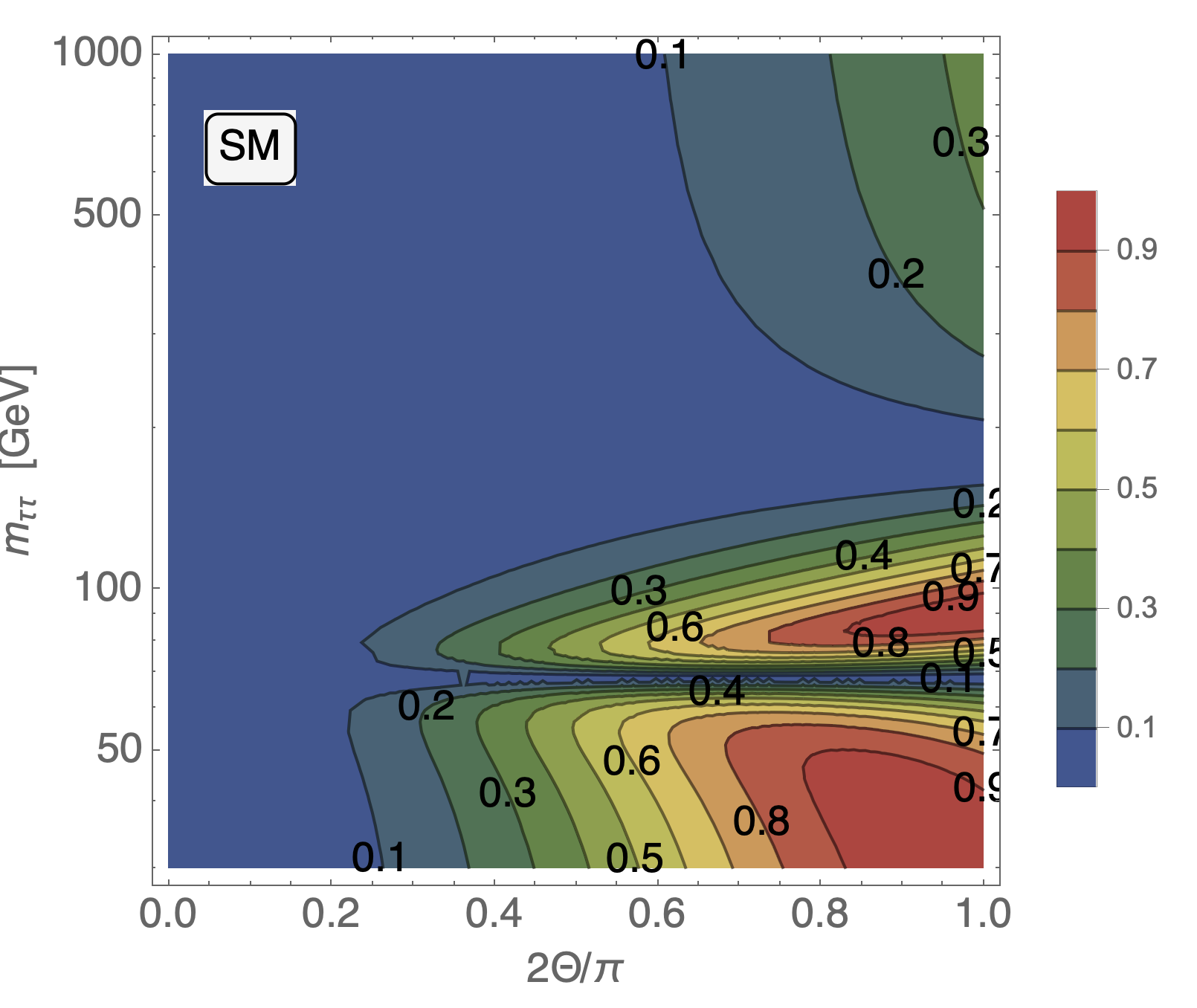}
\includegraphics[width=3.3in]{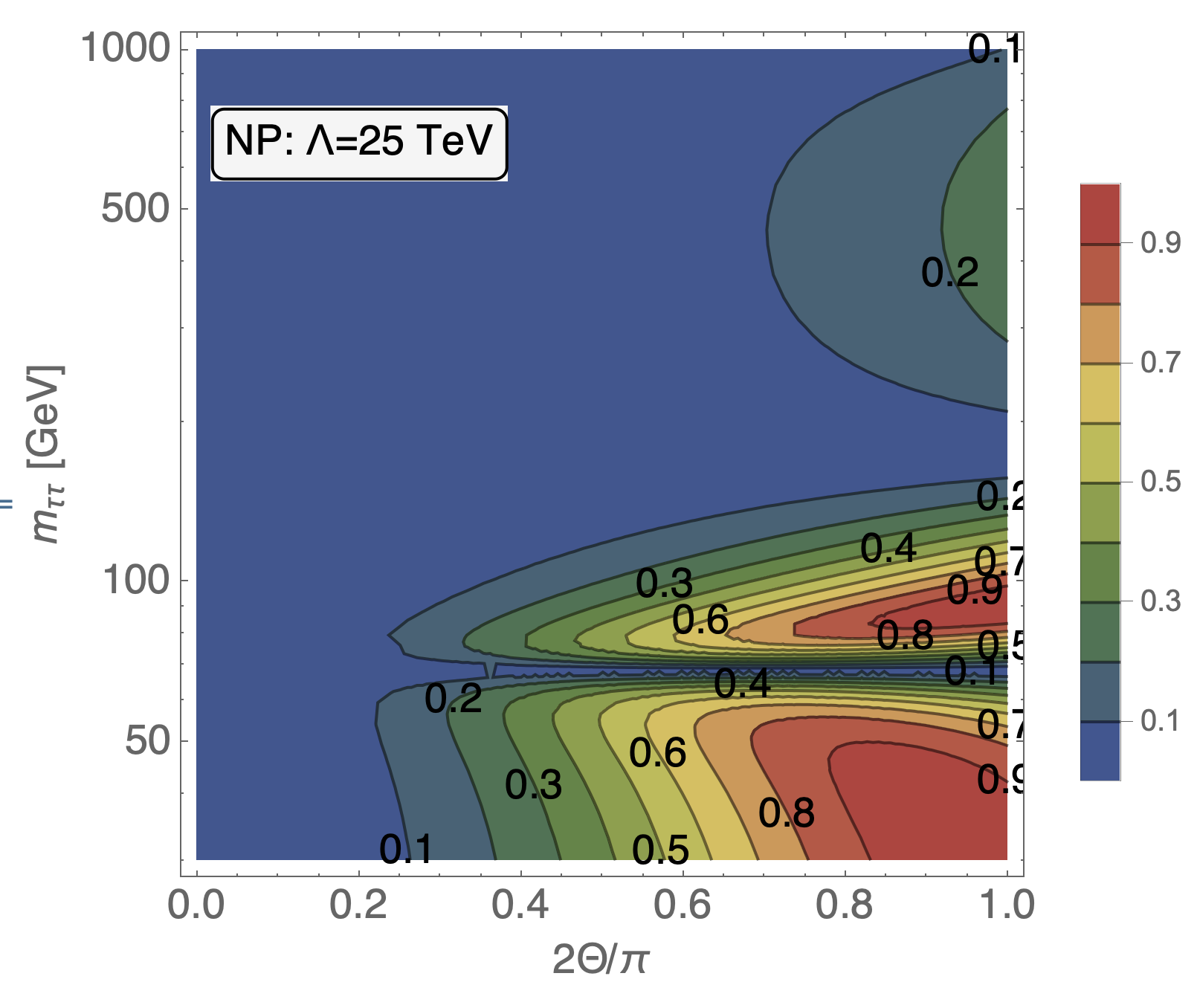}
\caption{\small \underline{On the left}: Concurrence ${\cal C}[\rho]$ in the SM as a function of the kinematical variables $\Theta$ and $m_{t \bar t}$. \underline{On the right}: Concurrence with new physics (NP) (with $\Lambda=25$ TeV). The invariant squared mass is on a logarithmic scale in both figures.
\label{fig:con-taus} 
}
\end{center}
\end{figure}
%%%%%%%%%%%%%%%%%%%%%% 

As we  already discussed, the entanglement becomes larger in the kinematical regions where either the photon or the $Z$-boson diagram dominates. Because the new physics terms increase as the energy in the CM, these  regions---being as they are at relatively low-energies---are not favorable for distinguishing between SM and new physics. It is at higher energies, just below 1 TeV that the two can best be compared. At these energies, the amount of entanglement is modest but very sensitive to the addition of new terms in the amplitude. We therefore consider the kinematical region $m_{\tau\bar\tau}>800$  as a compromise between having enough events and having new-physics effects sizable.

As previously mentioned, a new feature of the $\tau$-lepton case is the non-vanishing of some single polarization $B_i^\pm$ terms
in the spin density matrix (\ref{rho}), which were instead all zero in in the case of the top-quark pairs spin states. 
Their presence makes the extraction of the concurrence ${\cal C}[\rho]$ possible only numerically.

%%%%%%%%%%%%%%
 \begin{figure}[h!]
\begin{center}
\includegraphics[width=2.15in]{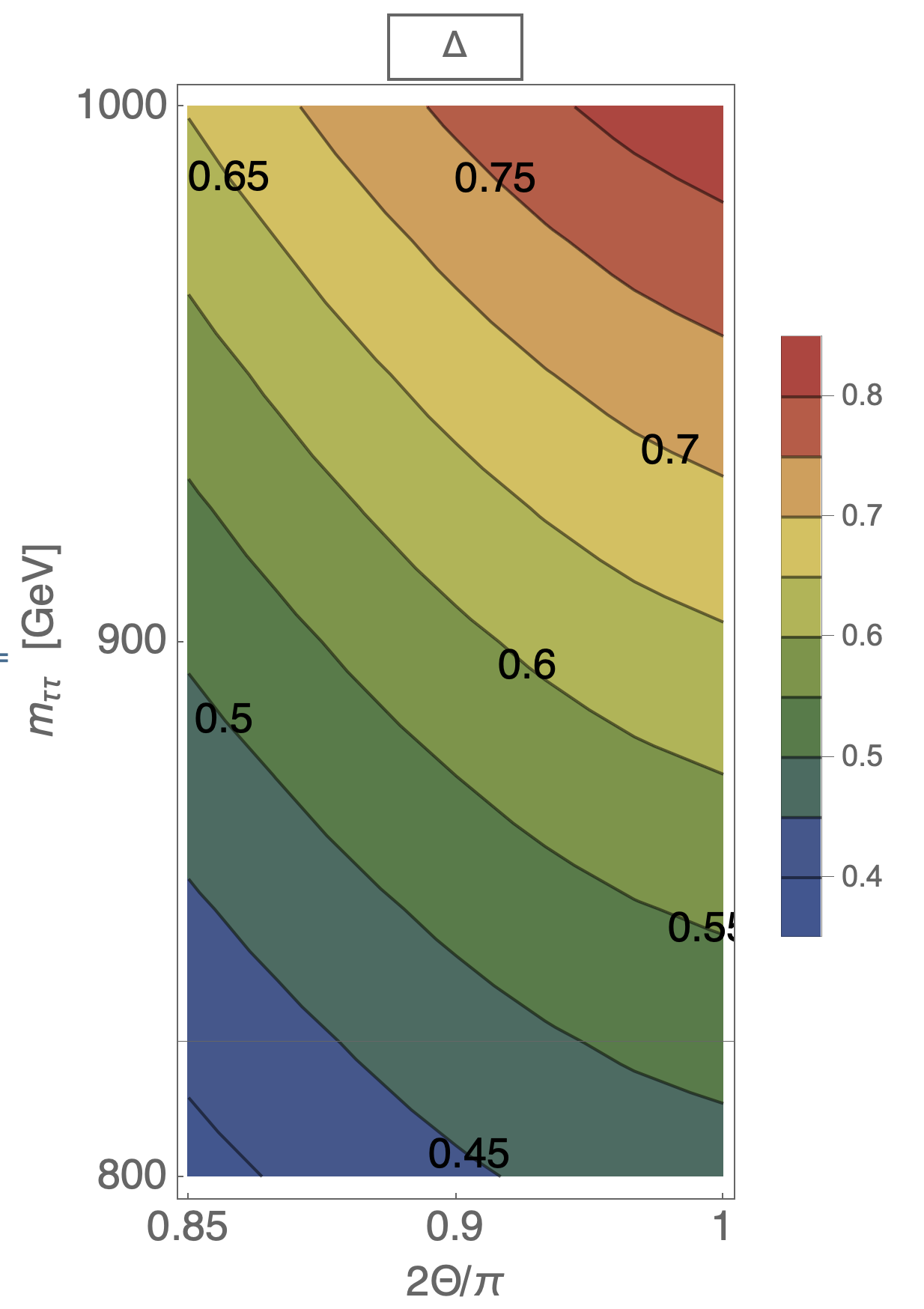}
\includegraphics[width=4in]{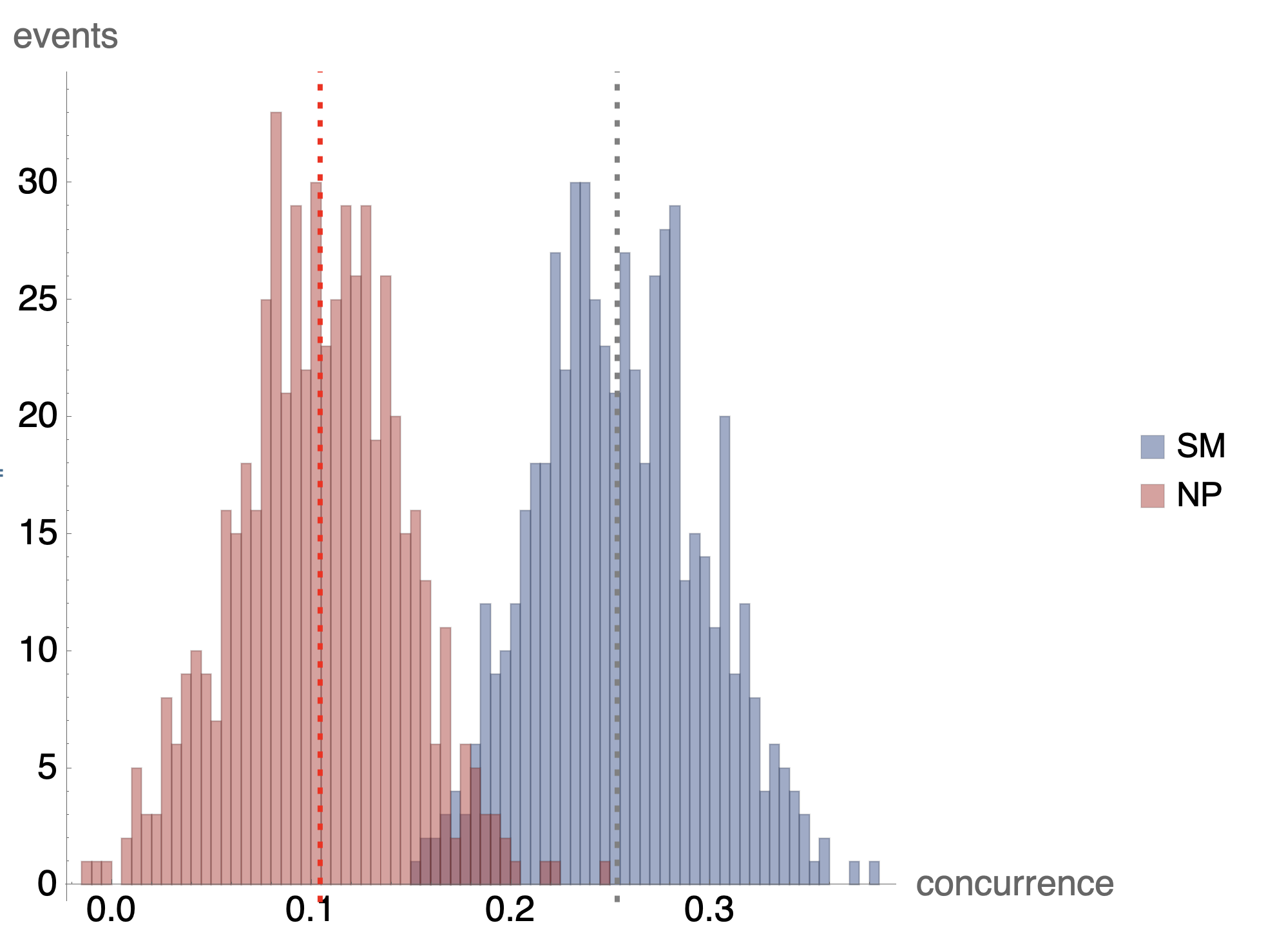}
\caption{\small \underline{On the left}: Percent difference in concurrence ${\cal C}[\rho]$ in the kinematical window $0.85 <2 \Theta/\pi < 1$ and $800 < m_{t\bar t} < 1000$ between   the SM and new physics  (with $\Lambda=25$ TeV or, equivalently, $c=0.02$) where $\Delta$ is defined as the difference between SM and new physics over the mean value for the SM. \underline{On the right}: Statistical distributions for the  573 events expected at Hi-Lumi for the SM (blue bins, mean value $0.25$) and the new physics (red bins, mean value $0.10$).
\label{fig:significanceDeltaTau} 
}
\end{center}
\end{figure}
%%%%%%%%%%%%%%%%%%%%%%

The left side of Fig.~\ref{fig:significanceDeltaTau} shows the kinematical region $m_{\tau\bar\tau}>800$ GeV and $0.85<x<1$ where the relative difference $\Delta$ between SM and new physics (with $\Lambda=25$ TeV) is largest and equal to about 70\%. The mean value of ${\cal C}[\rho]$ for the SM is $0.25$, $0.10$ for the new physics, the dispersion is about  $\sigma=0.04$. Such a large effect shows that  the contact interaction and its scrambling of the two $\tau$-lepton polarizations is a very effective way of changing the concurrence of their spins.

The number of events at high-energy is shown in Table~\ref{tab:events_taus}. They turn out to be too few at LHC run2. 
The right side of Fig.~\ref{fig:significanceDeltaTau} shows how  new physics can be distinguished from SM with  a significance of  3.7 for a contact interaction with a scale $\Lambda=25$ TeV with 573 events (Hi-Lumi). This result compares favorably with current determinations of four-fermion operators~\cite{ATLAS:2017eqx,CMS:2018ucw} (see also, the review in \cite{Workman:2022ynf}).

The value $\Lambda=25$ TeV corresponds to $c=0.02$ for $\Lambda=1\, \text{TeV}$ in the SMEFT notation where the contact operator  we are considering is given by
\be
\frac{c}{\Lambda^2}  ( \bar q_L \gamma^\alpha q_L) \,(\bar \tau_R \gamma_\alpha \tau_R ) \, .
\ee

This value is such as to make the estimate in a region where both the SMEFT expansion is safe and the NLO contributions smaller. Contrary to the magnetic dipole momentum in the top-quark case, in the $\tau$-lepton case EW and QCD NLO corrections are expected to be much smaller than the effect of the  contact interaction  terms.

\subsection{The  $\tau$-lepton decays}

The $\tau$ lepton decays as 
\bea
\tau^-  &\to&A + \nu_\tau \nn \\
& & \rotatebox[origin=c]{180}{$\looparrowleft$} \; A= e^-\bar \nu_e,\, \mu^- \bar \nu_\mu,\,\pi^- ,\, \rho^- ,\,a_1
\ldots
\eea
with
\be
\BR (\ell^- \bar\nu_\ell \nu_\tau) \simeq 35\%,\quad \BR (\pi^-  \nu_\tau) \simeq 11\%,  \quad \text{and} \quad \BR (\rho^-/a_1  \nu_\tau) \simeq 26\%\, .
\ee
These decays taken together account for more than 80\% of all decays.

The  angular distribution of the decay products can be used to determine the polarization of the $\tau$ lepton. The decay mode $\tau^-\to \pi^-  \nu_\tau$ has a distribution
\be
\frac{1}{\Gamma} \frac{\di \Gamma}{\di z} = \frac{1}{2} (1 + P_\tau z )\, , \label{pol0}
\ee
where $z=\cos \theta$---the angle being defined with respect to the charged final state. \eq{pol0} is the same as in the case of the top-quarks and  the same procedure to compute the coefficients $C_{ij}$  can be used here. Unfortunately the branching ratio for the  two $\tau$s to simultaneously decay in this channel is small and around 1\%.  Yet the large number of expected events around  $m_{\tau \bar \tau} \simeq 30$ GeV makes the analysis of the Bell inequalities violation possible.

The leptonic decay mode $\tau^-\to \ell^-  \bar \nu_\ell \nu_\tau$ has a distribution
\be
\frac{1}{\Gamma} \frac{\di \Gamma}{\di z } = \frac{1}{2} (1-z) \Big[(5+5z - 4 z^2) + P_\tau (1+z - 8z^2 )\Big]\, , \label{pol1}
\ee
which has only a weak dependence on the polarization~\cite{Bullock:1992yt}.

Therefore one has to resort to the two decays $\tau \to \rho\nu_\tau $ and $\tau\to a_1\nu_\tau$. The reconstruction of   the coefficients $C_{ij}$ for these channels depends on that of the polarizations of the mesons~\cite{Bullock:1992yt}. For this reason, the   analysis can only be done  by means of a full simulation. 
 The relevance of the physics of entanglement we discussed might encourage the  on-going efforts in this direction by the experimental collaborations. The effect is there---waiting to be extracted from the data.

\begin{figure}[h!]
\begin{center}
\includegraphics[width=3in]{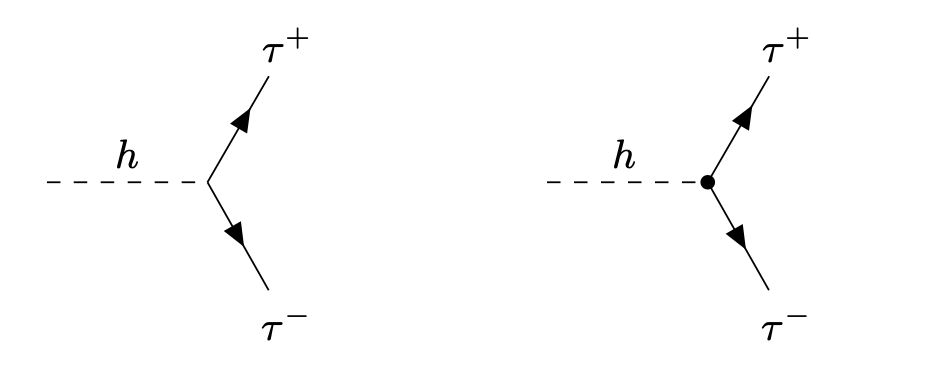}
\caption{\small \label{fig:taus_CP} Feynman diagrams for $\tau^-\tau^+$ production. \underline{On the right}, a possible CP odd vertex.}
\end{center}
\end{figure}
%%%%%%%%%%%%%%%%%%%%%%%%%%%%%%%%%%%%%%%%%%%%%

\subsection{Higgs boson decay}
The decay of the Higgs boson into a pair of fermions (see, Fig.~\ref{fig:taus_CP}) or---as we discuss in the next section, two photons---provides a physical process very similar to those utilized in atomic physics for studying entanglement. Because the final states originate from a scalar state, their entanglement is obvious as much as the correlation between their angular momenta.

\subsubsection{Bell inequalities} 
The interaction Lagrangian for the decay of the Higgs boson  into a pair of $\tau$ leptons is given by 
\be
{\cal L}_{\text{\tiny SM}} = \frac{m_\tau}{v} \bar \tau \tau \, h \, ,
\ee
where $v$ is the vacuum expectation value of the Higgs field $h$.
On the basis of this interaction term, the elements of the matrix $C_{ij}$ entering the tau lepton-pair spin density matrix (\ref{rho}) can be easily computed 
and given by
\be
C=  \begin{pmatrix} 1&0&0\\
0&1&0 \\
0& 0 & -1
\end{pmatrix} \, ,
 \ee
where the $C$ matrix above is defined on the $\{\hat{n},\hat{r},\hat{k}\}$ spin basis as in \eq{matrixC}.
 The sum of the square of the two largest eigenvalues gives $\mathfrak{m}_{12}[C]=2$, so that the 
Bell inequality (\ref{CHSH}) is maximally violated.

 %%%%%%%%%%%%
\begin{table}[h!]
\bc
\begin{tabular}{ccc}
%\toprule
&\hskip0.5cm (run 2) ${\color{oucrimsonred} {\cal L}=140\ \text{fb}^{-1}}$  \hskip0.5cm &  \hskip0.5cm (Hi-Lumi) ${\color{oucrimsonred} {\cal L}=3\ \text{ab}^{-1}}$  \hskip0.5cm \\[0.2cm]
\hline\\
 \underline{events} \hskip0.4cm  &\hskip0.4cm  $2.3 \times 10^5$  \hskip0.4cm &\hskip0.4cm  $5.1 \times 10^6$ \hskip0.4cm \\[0.4cm]
   \hline%
\end{tabular}
\caption{\label{tab:events_h_tau}Number of expected events in the Higgs boson decay into  $\tau^+\tau^-$ pairs.}
\ec
\end{table}
%%%%%%%%%%%%%%%%%%

\begin{figure}[h!]
\begin{center}
\includegraphics[width=3.5in]{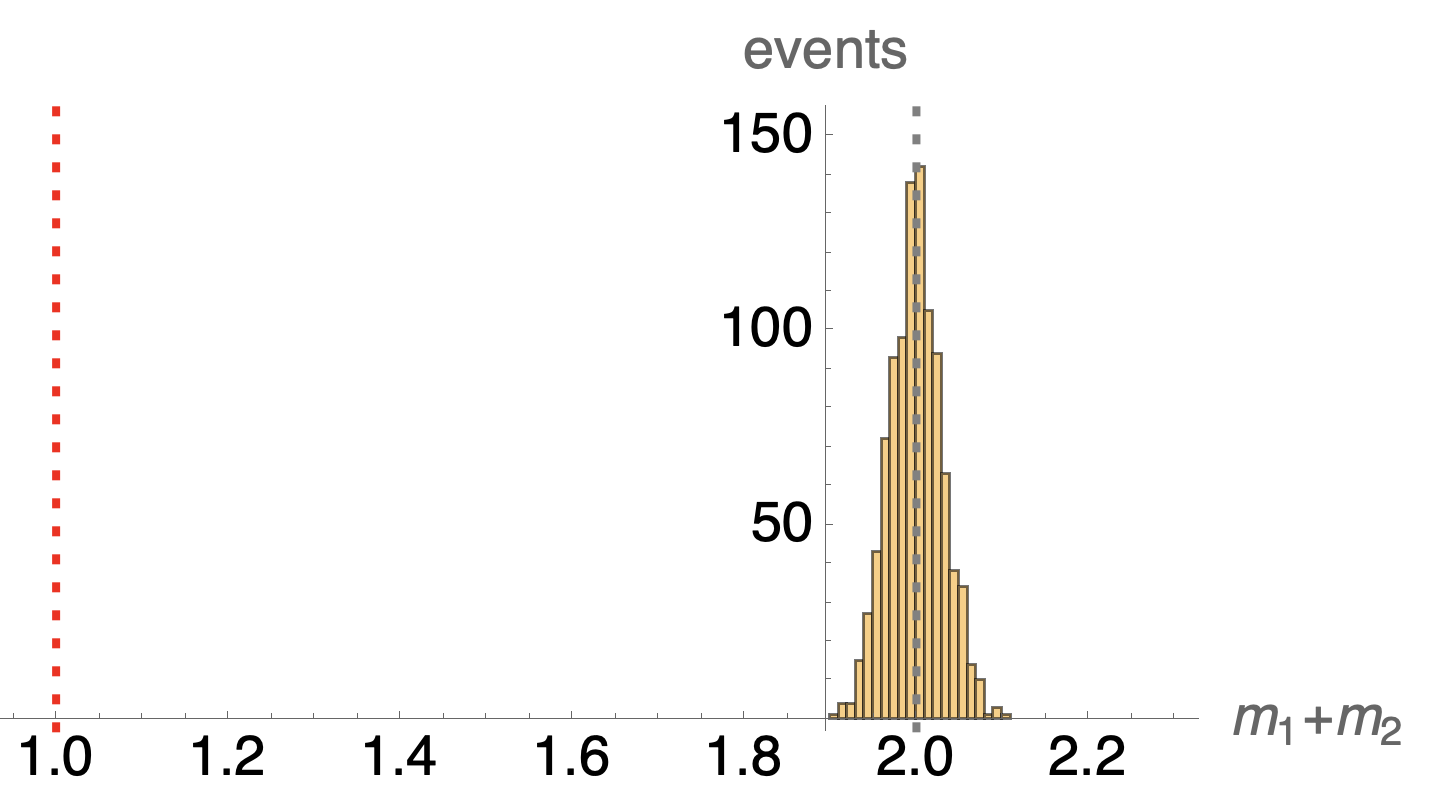}
\caption{\small Significance of the violation of Bell inequality in the decay of the Higgs boson in $\tau^-$-$\tau^+$  and photon pairs (as discussed below). Given the large number of events, the Gaussian distribution  around the value 2 is very peaked  ($\sigma=1/\sqrt{N}$) and the statistical significance large. As an example, we draw the case $N=1000$ for which the statistical significance of $\mathfrak{m}_{12}[C]>1$ is 32.
\label{fig:significanceHiggsTAU} 
}
\end{center}
\end{figure}

Since there is no kinematical dependence, we simply give the uncertainty as $1/\sqrt{N}$ with $N$ the number of events. Given the maximal violation of Bell inequalities and the large number of  events (see Table~\ref{tab:events_h_tau}   in which the cross sections are computed by running {\sc MADGRAPH5}~\cite{Alwall:2014hca} at the LO and then correcting by the $\kappa$-factor given at  the N3LO+N3LL~\cite{Bonvini:2016frm}), the significance can be huge. As shown in Fig~\ref{fig:significanceHiggsTAU} in the case of 1000 events, the significance of the hypothesis 
$\mathfrak{m}_{12}[C]>1$ is 10. Unfortunately in this kinematical region the tail of the pole of the $Z$-boson still dominates and gives a background that has to be reduced in order to proceed with the physical analysis.

%%%%%%%%%%%%%%
%%%%%%%%%%%%%%%%%%%%%% 

\subsubsection{Constraints on new physics}

We would like  to study the entanglement of the two $\tau$ leptons in the presence of new physics. The Higgs boson in the SM is a scalar CP even state. A CP odd component can be introduced  by a vertex as that in the Lagrangian~\cite{Desch:2003rw,Rouge:2005iy,Berge:2014sra}
\be
 {\cal L}_{\text{\tiny CP odd}}= i \frac{m_\tau}{v} \bar \tau \gamma_5 \tau\, h\, .
\ee
 For a recent review on possible CP odd interactions of the Higgs boson, see~\cite{Gritsan:2022php}.

Combining the two interactions, mimicking for instance the two doublet Higgs models, we have
\be
\cos \varphi \, {\cal L}_{\text{\tiny SM}}  + \sin \varphi \, {\cal L}_{\text{\tiny CP odd}} \, , \label{lagH}
\ee
where the parameter $\varphi$ modulates the amount of new physics.   

Taking the two vertices in \eq{lagH} together should provide the means to constrain the new-physics CP odd vertex. Absorptive contributions
are for the moment excluded, so that the coupling constants are assumed to be real.

Notice that, concerning the specific case of the on-shell Higgs decay in two fermions, the new physics parametrization given in \eq{lagH} is also the most general one. Indeed, in the case of both the Higgs and fermions fields on-shell, the  potential contribution of Lorentz invariant operators of higher dimensions can always be projected into the coupling structure provided by the renormalizable Lagrangian in \eq{lagH}.

With this generalized interaction Lagrangian, the elements of the correlation matrix $C$ in \eq{matrixC} become 
\be
C=  \begin{pmatrix} \displaystyle \frac{\beta_\tau^2 \cos^2\varphi - \sin^2\varphi}{\beta_\tau^2 \cos^2\varphi + \sin^2\varphi}&\quad \displaystyle -\frac{\beta_{\tau} \sin 2\varphi}{\beta_\tau^2 \cos^2\varphi + \sin^2\varphi}&\quad 0\\[0.5cm]
\displaystyle \frac{\beta_{\tau} \sin 2\varphi}{\beta_\tau^2 \cos^2\varphi + \sin^2\varphi}&\displaystyle \frac{\beta_\tau^2 \cos^2\varphi - \sin^2\varphi}{\beta_\tau^2 \cos^2\varphi + \sin^2\varphi}&0 \\[0.5cm]
0& 0 & -1
\end{pmatrix} \, ,
\ee
where now $\beta_\tau = \sqrt{1 - 4 m_\tau^2/m_h^2} $. This result has already been found in \cite{Bernreuther:1997gs,Desch:2003rw}. 
Then, the eigenvalues of the $M=C^TC$ matrix are equal to $(1,1,1)$ and the operator $\mathfrak{m}_{12}[C]=2$, showing a maximal violation of the Bell inequalities.

Similarly, the resulting concurrence is still maximal,
\be
{\cal C}[\rho]= 1 
\ee  
as it is independent from $\varphi$. This surprising result can be understood as follows.
At tree level, the interaction in (\ref{lagH}) produce pair of leptons that turn out to be totally unpolarized,
but highly correlated in spin. In fact, choosing the $z$-axis along the $\tau^-$ direction of flight in the Higgs rest frame,
and neglecting terms of order $(m_\tau/m_h)^2$, so that $\beta_\tau\simeq 1$,
the spin state of the $\tau$-lepton pair turns out to be \cite{Rouge:2005iy}:
\be
|\psi_{\tau\bar\tau}\rangle = \frac{1}{\sqrt{2}}\Big( |01\rangle + e^{2i\varphi}\, |10\rangle \Big)\ ,
\label{CP-state}
\ee
where
 $|0\rangle$ and $|1\rangle$ are as before the eigenvectors of $\sigma_z$, representing
the projection of the lepton spins along the $z$ axis.
As the $CP$ transformation reverses these spin projections, the pure state (\ref{CP-state}) is a $CP=1$ state for $\varphi=\,0$,
the usual SM result, while it is a $CP=-1$ state for $\varphi=\pi/2$.
In addition, it is maximally entangled for all values of $\varphi$: 
indeed, the trace over either leptons of the corresponding
density matrix gives a totally unpolarized state: 
${\rm Tr}_{1,2}\big[|\psi_{\tau\bar\tau}\rangle \langle\psi_{\tau\bar\tau}|\big]= \mathbb{1}/2$.
As a consequence, in this particular case, the entanglement content of the lepton pair spin state can not be used 
to bound $CP$-odd additions to the SM
as the spin quantum correlations are insensible to the angle $\varphi$.

A difference in entanglement shows in the concurrence  only in the presence of  an absorptive term---it  makes  the entanglement  no longer unaffected by the CP odd term. This result is reminiscence of what happens in the Kaon system where we need both a CP-violating  and a CP-conserving phase in order to be able to see direct  CP-violation. 

Such an  absorptive part can come in the SM from 
 QED loop correction to the vertex as the two final leptons exchange a photon but is very small. A phase could also be produced by the new physics term but we do not explore it further since is model dependent and it unavoidably introduces extra parameters and an uncertainty that is  hard to judge.

\section{Results: Two photons} 

%%%%%%%%%%%%%%%%%%%%%%%%%%%%%%%%%%
 \begin{figure}[h!]
\begin{center}
\com{%\tikzfeynmanset{ every vertex = {dot}}
  \feynmandiagram [small,horizontal=a to b] {
a -- [scalar, edge label=\(h\)] b,
f1 [particle=\(\gamma\)] -- [boson] b -- [boson] f2 [particle=\( \gamma\)],
};
\hspace{1.5cm}
  \feynmandiagram [small,horizontal=a to b] {
a -- [scalar, edge label=\(h\)] b [dot],
f1 [particle=\(\gamma\)] -- [boson] b -- [boson] f2 [particle=\( \gamma\)],
};}
\includegraphics[width=3in]{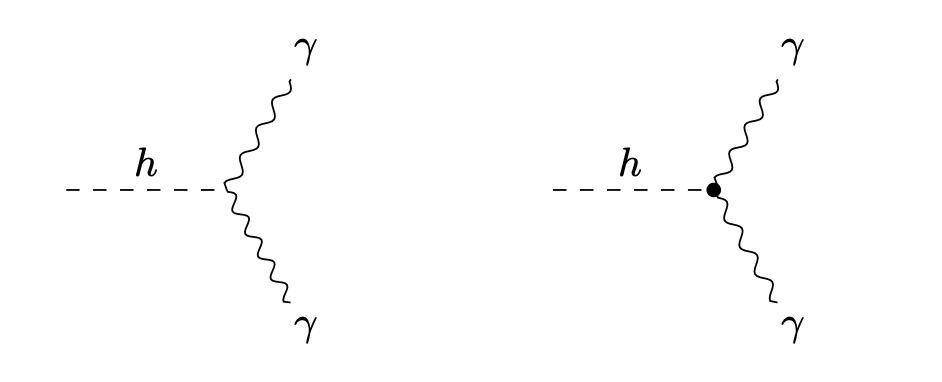}
\caption{\small \label{fig:gamma} Feynman diagrams for the Higgs boson $h$  into two photons. The dot stands for the CP-odd vertex.}
\end{center}
\end{figure}
%%%%%%%%%%%%%%%%%%%%%%%%%%%%%%%%%%%%%%%%%%%%%

The entanglement of a system of two photon has been discussed in~\cite{Lyuboshitz:2007zzb}. Here we examine it as the final states of the decay of the Higgs boson. In more than one way, this system is even closer than the Higgs boson decay into $\tau$-lepton pairs discussed in the previous section to what is done in atomic physics, where the polarizations of photons originating in atomic transitions are discussed.

\subsection{Bell inequalities}

The Higgs boson $h$ decays into two photons via an effective coupling $g_{\gamma\gamma h}$ provided in the SM by loop contributions. The Lagrangian is given by
\be
{\cal L} = -\frac{1}{4} \, g_{\gamma\gamma h} \, h \, F^{\mu\nu}F_{\mu\nu}\, ,
\ee
where $F^{\mu\nu}$ is the field strength of the photon.

The corresponding polarized amplitude square is 
\be
|{\cal M}_h|^2 = |g_{\gamma\gamma h}|^2 V^{\mu\nu}(k_1,k_2) V^{\rho\sigma}(k_1,k_2)
\left[\varepsilon^{\lambda_1}_\mu (k_1) \varepsilon^{\lambda^{\prime}_1 \ast}_\rho(k_1)\right] \left[\varepsilon^{\lambda_2}_\nu (k_2) \varepsilon^{\lambda^{\prime}_2 \ast}_\sigma (k_2)\right]\, ,
\ee
where $V^{\mu\nu}(k_1,k_2)=g^{\mu\nu}(k_1\cdot k_2)-k_1^{\nu}k_2^{\mu}$. Notice that, gauge invariance is guaranteed by the Ward Identities $k_1^{\mu}V_{\mu\nu}(k_1,k_2)=k_2^{\nu}V_{\mu\nu}(k_1,k_2)=0$.

The projection on the linear polarizations can be performed by substituting the terms in square brackets with the corresponding density matrix $\rho_{\mu\nu}$ given in \eq{densitymat}, and using the method explained in Section II.
After summing over all photons polarizations, we obtain the unpolarized amplitude square
\be
|\bar{{\cal M}}_h|^2=|g_{h\gamma\gamma}|^2 m_h^4/2\, ,
\label{M2hS}
\ee
to which corresponds the width $\Gamma=g_{h\gamma\gamma}^2 m_h^3/(64\pi^2)$.

After normalization over the unpolarized square amplitude in \eq{M2hS}, we find that   the correlation matrix $C$ is
 \be
C =  \begin{pmatrix} 1&0&0\\
0&-1&0 \\
0& 0 & 1
\end{pmatrix} \, . \label{mC}
\ee
in the basis of the Stokes parameters $\{\xi_1,\xi_2,\xi_3\}$ for the linear polarizations as defined in \eq{densitymat}. 
For the matrix $C$ in \eq{mC}, the operator $\mathfrak{m}_{12}[C]=2$  and the Bell inequalities are maximally violated. 
 
 %%%%%%%%%%%%
\begin{table}[h!]
\bc
\begin{tabular}{ccc}
%\toprule
&\hskip0.5cm (run 2) ${\color{oucrimsonred} {\cal L}=140\ \text{fb}^{-1}}$  \hskip0.5cm &  \hskip0.5cm (Hi-Lumi) ${\color{oucrimsonred} {\cal L}=3\ \text{ab}^{-1}}$  \hskip0.5cm \\[0.2cm]
\hline\\
 \underline{events} \hskip0.4cm  &\hskip0.4cm   $8854$  \hskip0.4cm &\hskip0.4cm  $1.8 \times 10^5$ \hskip0.4cm \\[0.4cm]
   \hline%
\end{tabular}
\caption{\label{tab:events_h_gamma} Number of expected events in the Higgs boson decay into  two photons.}
\ec
\end{table}
%%%%%%%%%%%%%%%%%%

To estimate the uncertainty, we consider a Gaussian distribution around the value 2.  
The number of events expected in the production and decay of the Higgs boson into two photons is shown in Table~\ref{tab:events_h_gamma}  in which the cross sections are estimated at the LO by means of {\sc MADGRAPH5}~\cite{Alwall:2014hca} and then corrected by the $\kappa$-factor as estimated at the
N3LO+N3LL~\cite{Bonvini:2016frm}. Given the large number of events, this distribution has a rather sharp peak.
The statistical significance of the violation is shown in Fig.~\ref{fig:significanceHiggsTAU} (Section 4.7) for 1000 events, the same way we did for the case of the Higgs boson decaying into a pair of $\tau$-leptons. Already for this number of events the violation is statistically significant. This a test well worth doing but it requires the detection of the polarization of the two photons.

\subsection{New physics}

As discussed in the case of the decay into $\tau$-leptons, the Higgs boson could have a CP odd vertex from a new physics contribution. In the case of two final photons, it would decay just like the neutral pion $\pi^0$ decays into two photons via the anomaly. Then, it is useful to parametrize the corresponding effective Lagrangian as 
\be
{\cal L}^\prime = - \frac{1}{4} \, \tilde{g}_{h\gamma\gamma}\, h \, F^{\mu\nu}\tilde F_{\mu\nu} \, ,
\ee 
where 
$\tilde{F}_{\mu\nu}=1/2 \epsilon_{\mu\nu\alpha\beta} F^{\alpha\beta}$ is the dual field strength of the photon, with $\epsilon_{\mu\nu\alpha\beta}$ the Levi-Civita
antisymmetric tensor satisfying $\epsilon_{0123}=1$.
Then, the corresponding polarized square amplitude is given by
\be
|{\cal M}^\prime_{h}|^2 =
 |\tilde{g}_{\gamma\gamma h}|^2 \tilde{V}^{\mu\nu}(k_1,k_2) \tilde{V}^{\rho\sigma}(k_1,k_2) \left[\varepsilon^{\lambda_1}_\mu (k_1) \varepsilon^{\lambda^{\prime}_1 \ast}_\rho(k_1)\right] \left[\varepsilon^{\lambda_2}_\nu (k_2) \varepsilon^{\lambda^{\prime}_2 \ast}_\sigma (k_2)\right]\, ,
\ee
where $\tilde{V}_{\mu\nu}(k_1,k_2)=\epsilon_{\mu\nu\alpha\beta}k_1^{\alpha}k_2^{\beta}$. Notice that, gauge invariance of the amplitude is automatically guaranteed by the antisymmetric properties of $\epsilon_{\mu\nu\alpha\beta}$.
For the unpolarized amplitude square, we obtain
\be
|\bar{{\cal M}}_h|^2=|\tilde{g}_{h\gamma\gamma}|^2 m_h^4/2\, .
\label{M2hS}
\ee

If we now combine the two Lagrangians as
\be
   {\cal L}^{\rm total}=\frac{g_{\gamma\gamma h}}{4} \left[
     \big( h \, F^{\mu\nu}F_{\mu\nu}\big) + \, z \big( h \, F^{\mu\nu}\tilde F_{\mu\nu}\big)\right]
   \label{Lhgg}
\ee
by collecting the respective coefficients in the parameter
$z\equiv \tilde{g}_{\gamma\gamma h}/g_{\gamma\gamma h}$, we might expect to be able to study the effect of the new physics on the entanglement of the two final photons. 

A non-vanishing interference among the two contributions arises in the polarized contributions to the square amplitude generated by the Lagrangian
in \eq{Lhgg}, while it vanishes in the unpolarized case.

From the Lagrangian in \eq{Lhgg} we obtain the correlation matrix  (in the Stokes parameter basis)
 \be
 C =  \begin{pmatrix} \displaystyle{\frac{1-z^2}{1+z^2}}&\quad 0&\quad
  \displaystyle{\frac{2\, z}{1+z^2}}\\[0.5cm]
0&-1&0 \\[0.25cm]
-\displaystyle{\frac{2\, z}{1+z^2}}& 0 &
\displaystyle{\frac{1-z^2}{1+z^2}}
\end{pmatrix} \, .
\ee
 
Non-vanishing contributions to the $B_i$ coefficients arise due to the interference term; they are given by
\be
B^+_2=-B_2^{-}= i\frac{2 |z|}{1+|z|^2}\, , ~  B^{\pm}_1=B^{\pm}_3=0.
\ee
They represents the polarization of the single photons.
The $B_i$ coefficients are purely imaginary and cancel out in the matrix $R$ (see \eq{rR} in Section II) which gives us the concurrence. 

As in the case of the final two $\tau$ leptons, for real coefficients in the Lagrangians in \eq{Lhgg}, our expectation of extracting bounds on the new physics is frustrated and the overall entanglement is not affected and 
\be
{\cal C}[\rho] = 1 \, .\label{cc-ph} 
\ee
even in the presence of new physics.

The presence of an absorptive part in one or both the two couplings 
$g_{\gamma\gamma h}$ and $\tilde{g}_{\gamma\gamma h}$, by inducing a phase
$\delta=\arg z$, would make the concurrence sensitive to the new physics. In this case, the matrix $C$ would become
  \be
 C =  \begin{pmatrix} \displaystyle{\frac{1-|z|^2}{1+|z|^2}}&\quad 0&\quad
  \displaystyle{\frac{2 |z|\cos \delta}{1+|z|^2}}\\[0.5cm]
0&-1&0 \\[0.25cm]
-\displaystyle{\frac{2 |z|\cos \delta}{1+|z|^2}}& 0 &
\displaystyle{\frac{1-|z|^2}{1+|z|^2}}
\end{pmatrix} \, ,
\ee
and (for $z$ and $\delta$ small) the concurrence is given by
\be
{\cal C}[\rho] = 1 -  |z| \, \delta \, ,\label{cc-ph2} 
\ee
thus probing the presence of the CP-odd term.

 A phase is  present in the SM vertex and  comes mainly from the absorptive part of the $b$-quark loop in the effective coupling between the Higgs boson and the two photons. 
In particular, assuming the  $\tilde{g}_{\gamma\gamma h}$ real, for the phase $\delta$ in the SM framework we get $\delta \sim -4\times 10^{-4}$.
Even if included,  this contribution is too small to make a difference.

One may speculate about the presence of a phase in the  CP odd vertex. Since the size of it depends on the specific model and the uncertainty would be hard to gauge, we do not to pursue this possibility further.

\subsection{Detecting photon polarizations}

The possibility of measuring photon polarizations  depends on their energy. For high-energy photons, the dominant process is pair production as the photons fly through matter. There are two possible processes: the electron interacting with the nuclei ($A$) or the atom electrons:
\bea
\gamma + A &\to& A+ e^+ + e^- \nn \\
\gamma + e^- &\to& e^-+ e^+ + e^-  \, ,
\eea
with the latter dominating in the energy range we are interested in.

For a polarized photon,  the Bethe-Heitler  cross section for the \textit{Bremsstrahlung} production of electron pairs depends also on the azimuthal angles $\varphi_\pm$ of the produced electron and positron~\cite{May,Borsellino:1953zz} as
\be
\frac{\di s}{\di \varphi_+ \di \varphi_- } = \sigma_0 \Big[  X_{\text{un}} + X_{\text{pol}} \, P_\gamma \cos (\varphi_+-\varphi_-) \Big] \, ,\label{pairs}
\ee
where $P_\gamma$ is the linear polarization fraction of the incident photon, $ X_{\text{un}}$ and   $X_{\text{pol}} $  are the unpolarized and  polarized coefficients respectively, which depend on the kinematical variables. The explicit form of the cross section in \eq{pairs} can be found in~\cite{Boldyshev:1994bs}. The relevant azimuthal information  comes from  the dependence of the cross section on the
a-coplanarity of the outgoing electron and positron.
The measure of the relative angle between these momenta gives information on the polarization of the photon.

Even though this possibility is not currently  implemented at the LHC, detectors able to perform such a measurement are under discussion for astrophysical $\gamma$ rays~\cite{Depaola:2009zz} and     an event generator to simulate the process already exists~\cite{Bernard:2013jea} and has been implemented within GEANT~\cite{GEANT4:2002zbu}
(for a recent review, see~\cite{Bernard}).

\section{Summary and outlook} 

 Our exploration of the use of quantum entanglement at colliders shows that it can provide new tests of quantum mechanics as well as a very promising new tool in the study of the SM as well as its possible extensions leading to new physics.
 
  The best system by far where to test the violation of Bell inequalities is the decay of the Higgs boson. The  decay of the Higgs  boson into a pair of $\tau$-leptons seems already feasible with the data of run2 at the LHC although one has to disentangle this channel from the tail of the $Z$-boson pole that still dominates at values of the invariant mass around the Higgs boson mass.  The  decay of the Higgs  boson into two photons, while equally promising, requires a dedicated detector in order to measure the photon polarizations. A similar test is also possible in the production of top-quark and $\tau$-lepton pairs, with the latter the most promising within the kinematical windows just above threshold and around the $Z$-boson pole.
 
 Searching for constraints to new physics is possible by means of the entanglement between the spins of pairs of particles produced in the collisions. We find the concurrence ${\cal C}[\rho]$ an observable very sensitive to any physics beyond the SM. The use of this tool is  illustrated  by considering a magnetic dipole operator in the case of having top-quark pairs  and a  contact interaction  for the case of $\tau$ leptons. Because the impact of the higher order operators corresponding to this new physics grows with the energy of the process, they must be tested at the highest energy available. In this regime, the top-quark pairs are mostly entangled while  the $\tau$ leptons less so. Nevertheless, what counts is the relative change in entanglement and both cases show a promising power in constraining the size of the new physics better than current limits based on total cross sections or classical correlations.

Our analysis is about pseudo-observables at the level of the parton production processes and decays. The last word---comprising a full analysis of all uncertainties, statistical as well as systematic---can only come from the extraction of the entanglement observables $\mathfrak{m}_{12}[C]$ and ${\cal C}[\rho]$  from the data (or at least from a full simulation of them). Though this is a challenging problem---which can only be properly addressed by the experimental collaborations---it is well worth the effort because of the improved sensitivity to which the  new tools  give access.

\section*{Acknowledgments}
         {\small
MF thanks M.\ Casarsa for help on the implementation of the toy Monte Carlo and L.\ Marzola for help on the PDFs. We thank M.\ Kadastik, L.\ Marzola, J.\ Pata and C.\ Veelken for discussions on the physics of the $\tau$ leptons. We thank M.\ Casarsa and F.\ Longo for discussions on the detection of photon polarizations. }

\appendix
%\raggedbottom

\section{Cross section functions}

In this Appendix we provide the explicit expression for the functions $\tilde A$, $\tilde C_{ij}$ and $\tilde B_i$ in the cross section of the processes discussed in the main text. The tilde above these quantities reminds us that they have to be normalized, like in \eq{pdf} and \eq{pdf2}, dividing by the unpolarized cross section  to give the  entries of \eq{rho}.

\subsection{Top-quark pairs}
\subsubsection{ Standard Model}

Below, we write the $\tilde{A}^{q\bar{q}}$, $\tilde{B}^{qq}_{i}$, and $\tilde{C}^{qq}_{ij}$  coefficients for the $t\bar t$ pair production via $q\bar{q}$ and $g g$ scattering in the SM framework.
These functions were calculated in~\cite{Bernreuther:1993hq,Uwer:2004vp}; we report them for ease of reading:

\begin{subequations}
\begin{align}
  \tilde{A}^{gg} &= F_{gg} \Big[1 + 2\beta_t^2\sin^2\Theta
  -\beta_t^4\left(1+\sin^4\Theta\right)\Big],\\
  \tilde{C}^{gg}_{nn} &= - F_{gg} \Big[1-2\beta_t^2
  +\beta_t^4\left(1+\sin^4 \Theta\right)\Big],\\
  \tilde{C}^{gg}_{rr} &=  - F_{gg} \Big[1
  - \beta_t^2 \left(2-\beta_t^2\right)\left(1+ \sin^4\Theta \right)  \Big],\\
    \tilde{C}^{gg}_{kk} &= - F_{gg} \Big[1 - \beta_t^2\frac{\sin^2 2\Theta}{2} -\beta_t^4\left(1+ \sin^4\Theta \right) \Big],\\
    \tilde{C}^{gg}_{kr} &=  \tilde{C}^{gg}_{rk}= F_{gg} \, \beta_t^2\sqrt{1-\beta_t^2} \sin 2\Theta\sin^2\Theta\, 
    \\   \tilde{B}^{gg}_k&=\; \tilde{B}^{gg}_r=\tilde{B}^{gg}_n=0\, ,
\end{align}
\end{subequations}
with $\displaystyle F_{gg} =  \frac{N_c^2\left(1+\beta_t^2\cos^2\Theta\right)-2}{64N_c\left(1-\beta_t^2 \cos^2\Theta\right)^2}$\, and
\begin{subequations}
\begin{align}
    \tilde{A}^{q\bar{q}} & = F_{q\bar{q}} \Big(2-\beta_t^2\sin^2\Theta\Big),\\
    \tilde{C}^{q\bar{q}}_{nn} &= -F_{q\bar{q}}\, \beta_t^2 \sin^2\Theta,\\
    \tilde{C}^{q\bar{q}}_{rr} &= F_{q\bar{q}} \Big(2-\beta_t^2 \Big)\sin^2\Theta,\\
    \tilde{C}^{q\bar{q}}_{kk} &= F_{q\bar{q}} \Big(2\cos^2\Theta+\beta_t^2 \sin^2\Theta \Big),\\
    \tilde{C}^{q\bar{q}}_{kr} &=\tilde{C}^{q\bar{q}}_{rk}= F_{q\bar{q}}\, \sqrt{1-\beta_t^2}\sin 2\Theta,
    \\   \tilde{B}^{gg}_k& =\; \tilde{B}^{gg}_r=\tilde{B}^{gg}_n=0\, ,
\end{align}
\end{subequations}
with $\displaystyle F_{q\bar{q}} = \frac{1}{2N^2_c}$.

\subsubsection{New physics: magnetic moment dipole}
Here, we collect the results  for the coefficients $\tilde{A}^{q\bar{q}}$, $\tilde{B}^{qq}_{i}$, and $\tilde{C}^{qq}_{ij}$  for the magnetic moment dipole in \eq{dipole} in which  we retained only the interference with the corresponding SM amplitudes. They were computed in~\cite{Bernreuther:2015yna}:

\begin{subequations}
	\begin{align}
          \tilde{A}^{gg}  = &\;  f_{gg}^{(1)} \, 
         \Big[ N^2_c \left(1+\beta_t^2 \cos^2\Theta\right) -2\Big] \; \mu,\\
%%%%%%%%%%%%%%%%%%%%%%%%%%%
  \tilde{C}^{gg}_{nn}  =&\;  f_{gg}^{(1)}\,
  \left(2-N^2_c\right) \mu,\\ 
  \tilde{C}^{gg}_{rr} = & \; f_{gg}^{(2)}\,
 \,  \Big[
	    N_c^2 \left(-1+\beta_t^4 \cos^2\Theta \sin^4\Theta \right)+ (N_c^2-2)\beta_t^2 \bigl(\sin^2\Theta+\cos^4\Theta\bigr)+ 2\Big] \mu, \\
%%%%%%%%%%%%%%%%%%%%%%%%%%%
 \tilde{C}^{gg}_{kk} =&\;   f_{gg}^{(2)}\,
 \frac{1}
       {1-\beta_t^2} \Bigg\{N_c^2 \Big[-1-\beta_t^2\bigl(-2+\cos^4\Theta\bigr)
    + \beta_t^6 \cos^2\Theta \bigl(-1-\cos^2\Theta+\cos^4\Theta\bigr) \nonumber \\
	 & \;-
    \beta_t^4 \bigl(\sin^2\Theta-2 \cos^4\Theta + \cos^6\Theta\bigr)\Big]
    \nonumber \\
	 & \; +  2\Big[1+\beta_t^4\bigl(1+\cos^2\Theta - \cos^4\Theta\bigr) + \beta_t^2 \bigl(-2-\cos^2\Theta + \cos^4\Theta \bigr)\Big]\Bigg\} \mu, \\
%%%%%%%%%%%%%%%%%%%%%%%%%%%
    \tilde{C}^{gg}_{kr}= &\; \tilde{C}^{gg}_{rk}= f_{gg}^{(2)}\,  
   \,  \frac{ \beta_t^2\sin 2\Theta}{2\sqrt{1-\beta_t^2}}\,\Bigg\{\frac{N_c^2}{2}\Big[-2 \cos^2\Theta
             - \beta_t^4 \cos^2\Theta \sin^2\Theta + \beta_t^2 \bigl(1+3 \cos^2\Theta - 2 \cos^4\Theta\bigr)\Big]\nn \\
          & -1 + \left(2-\beta_t^2\right)\cos^2\Theta\Bigg\} \mu \, ,
 \\          \tilde{B}^{gg}_k=&\; \tilde{B}^{gg}_r=\tilde{B}^{gg}_n=0\, ,
	\end{align}
	\end{subequations}
	and 
	\begin{subequations}
     \begin{align}
	   \tilde{A}^{q\bar{q}}  = & \; \frac{f_{q\bar q}}{2}\;\mu, \\
	  \tilde{C}^{q\bar{q}}_{nn}= &\; 0, \\
          \tilde{C}^{q\bar{q}}_{rr} = &\;
          \frac{f_{q\bar q}}{2}\left(1-\cos^2\Theta\right)\; \mu , \\
	 \tilde{C}^{q\bar{q}}_{kk} = &\; \frac{f_{q\bar q}}{2}\, \cos^2\Theta \;\mu, \\
	  \tilde{C}^{q\bar{q}}_{kr} = &\; \tilde{C}^{q\bar{q}}_{rk} = f_{q\bar q}\,\frac{\left(2-\beta_t^2\right)\sin2\Theta }{8\sqrt{1-\beta_t^2}}\; \mu,
          \\       \tilde{B}^{q\bar{q}}_k=&\; \tilde{B}^{q\bar{q}}_r=\tilde{B}^{q\bar{q}}_n=0\, ,
	\end{align}
        \end{subequations}
        with $\displaystyle f_{gg}^{(1)}=\frac{1}{N_c (N_c^2-1)}\frac{1}{(1-\beta_t^2\cos^2 \Theta)}$, $\displaystyle f_{gg}^{(2)}=\frac{1}{N_c (N_c^2-1)}\frac{1}{(1-\beta_t^2\cos^2 \Theta)^2}$ and $\displaystyle f_{q\bar q}=\frac{N_c^2-1}{N_c^2}$.
%\clearpage

\subsection{$\tau$-lepton pairs}
 
\subsubsection{Standard Model}
Below, we write the $\tilde{A}^{q\bar{q}}$, $\tilde{B}^{qq}_{i}$, and $\tilde{C}^{qq}_{ij}$  coefficients for the $\tau^+\tau^-$ pair production via $q\bar{q}$ scattering in the SM framework:
\begin{subequations}
  \begin{align}
%%%%%%%%%%%%%%%%%%%%%%%%%%
  \tilde{A}^{q\bar{q}} &= F_{q\bar{q}}\bigg\{
  Q_q^2 Q_\tau^2 \Big[2-\beta_\tau^2 \sin^2\Theta \Big] + 2 Q_q Q_\tau
  \Rechi \Big[ 2 \beta_\tau g_A^q g_A^\tau\cos\Theta + g_V^q g_V^\tau\left( 2  - \beta_\tau^2 \sin^2\Theta\right) \Big]
  \nn \\
  & + \Abschi^2 \bigg[ \Big(g_V^{q 2}+ g_A^{q 2}\Big)\Big(2g_V^{\tau 2}+
      2\beta_\tau^2 g_A^{\tau 2}-\beta_\tau^2  \left(g_V^{\tau  2}+ g_A^{\tau  2}\right)\sin^2\Theta
      \Big)
  + 8 \beta_\tau  g_V^q g_V^\tau g_A^q g_A^\tau  \cos \Theta
  \bigg]\bigg\}\, , \\
\nn\\
%%%%%%%%%%%%%%%%%%%%%%%%%%
\tilde{C}^{q\bar{q}}_{nn} &= -F_{q\bar{q}} \beta_\tau^2 \sin^2\Theta
\bigg\{ Q_q^2 Q_\tau^2 + 2 Q_q Q_\tau\Rechi g_V^{q} g_V^{\tau} -\Abschi^2\Big(g_V^{q 2}+g_A^{q 2}\Big) 
\Big(g_A^{\tau 2}-g_V^{\tau 2}\Big)\bigg\}\, , \\
%%%%%%%%%%%%%%%%%%%%%%%%%%%%%%%%%%%%%%%%%%%%%%%%%%%
\tilde{C}^{q\bar{q}}_{rr} &= -F_{q\bar{q}} \sin^2\Theta \,\bigg\{ \left(\beta_\tau^2-2\right)     Q_q^2 Q_\tau^2  + 2 Q_q Q_\tau \Rechi g_V^q g_V^\tau \left(\beta_\tau^2-2\right) \nn \\
& +
    \Abschi^2  \Big[\beta_\tau^2 \left(g_A^{\tau 2}+g_V^{\tau 2}\right) - 2g_V^{\tau 2}\Big] \left(g_V^{q 2}+ g_A^{q 2}\right) \bigg\}\, ,
    \\
%%%%%%%%%%%%%%%%%%%%%%%%%%  
\tilde{C}^{q\bar{q}}_{kk} &= F_{q\bar{q}}\bigg\{ Q_q^2 Q_\tau^2 \Big[ \left(\beta_\tau^2-2\right)  \sin^2\Theta + 2 \Big] \nn \\
& + 2 Q_q Q_\tau \Rechi\Big[2 \beta_\tau  g_A^q g_A^\tau  \cos \Theta  + g_V^q g_V^\tau \big( (\beta_\tau^2 -2) \sin^2\Theta + 2 \big) \Big]
\nn \\
& + \Abschi^2 \Big[8 \beta_\tau g_A^q g_A^\tau g_V^q g_V^\tau \cos \Theta
  +\Big(g_V^{q 2}+g_A^{q 2}\Big)\Big(
  2g_V^{\tau 2} \cos^2\Theta
  - \beta_\tau^2 \left(g_A^{\tau 2}-g_V^{\tau 2}\right)\sin^2\Theta
  +2\beta_\tau^2g_A^{\tau 2}\Big)
\Big]\bigg\}\, , \\
%%%%%%%%%%%%%%%%%%%%%%%%%%%%%%%%%%%%%%%%%%%%%%%%%%%        
    \tilde{C}^{q\bar{q}}_{kr} &= \tilde{C}^{q\bar{q}}_{rk} = 2 F_{q\bar{q}}  \sin \Theta \sqrt{1-\beta_\tau^2} \Bigg\{ Q_q^2 Q_\tau^2 \cos\Theta + 
     Q_q Q_\tau   \,\Rechi \Big[ \beta_\tau g_A^q g_A^\tau + 2 
     g_V^q g_V^\tau\cos \Theta \Big] \nn \\ 
     & + \Abschi^2 \Big[ 2\beta_\tau  g_A^q g_A^\tau g_V^q g_V^\tau
     +  g_V^{\tau 2} \left(g_V^{q 2}+ g_A^{q 2}\right) \cos \Theta  \Big] \Bigg\}\, , \nn \\
 %%%%%%%%%%%%%%%%%%%%%%%%%%%%%%%%%%%%%%%%%%%%%%%%%%%%%%%%%%%%%%%%%
     \tilde{C}^{q\bar{q}}_{rn} &=  \tilde{C}^{q\bar{q}}_{nr}=
     \tilde{C}^{q\bar{q}}_{kn}=\tilde{C}^{q\bar{q}}_{nk}=0\, ,
%%%%%%%%%%%%%%%%%%%%%%%%%%%%%%%%%%%%%%%%%%%%%%%%%%%        
     \\ \nn \\
     \tilde{B}^{q\bar{q}}_{k} &= -2F_{q\bar{q}}\bigg\{   
     Q_qQ_\tau\Rechi\Big[\beta_\tau g_A^{\tau}g_V^{q}\left(1+\cos^2\Theta\right)
       +2 g_A^{q}g_V^{\tau}\cos\Theta\Big] 
     \nn \\
&+ \Abschi^2 \left[2 g_A^{q}g_V^{q}\Big( \beta_\tau^2g_A^{\tau 2} +g_V^{\tau 2}\Big)\cos\Theta+
       \beta_\tau g_A^{\tau}g_V^{\tau}\left(g_V^{q 2}+g_A^{q 2}\right)\left(1+\cos^2\Theta\right)\right]
     \bigg\}\, ,
    \\ 
%%%%%%%%%%%%%%%%%%%%%%%%%%%%%%%%%%%%%%%%%%%%%%%%%%%        
    \tilde{B}^{q\bar{q}}_{r} &= -2F_{q\bar{q}}\sin\Theta\sqrt{1-\beta_\tau^2}\bigg\{Q_qQ_\tau\Rechi\Big[\beta_\tau  g_A^{\tau}g_V^{q}\cos\Theta
      +2g_A^{q}g_V^{\tau}\Big] \nn \\
    &+ \Abschi^2 g_V^{\tau}\Big[\beta_\tau g_A^{\tau}
      \left(g_V^{q 2}+g_A^{q 2}\right)\cos\Theta +2 g_A^q g_V^{q}g_V^{\tau}\Big]     \bigg\}\, ,
    \\
%\nn\\
%%%%%%%%%%%%%%%%%%%%%%%%%%%%%%%%%%%%%%%%%%%%%%%%%%%        
      \tilde{B}^{q\bar{q}}_{n} &= 0\, ,
\end{align}
\end{subequations} 
with $\displaystyle F_{q\bar{q}} =  \frac{1}{16}$, $Q_{q,\tau}$ the electric charges, $\beta_\tau$ the 
$\tau^{\pm}$ velocity in their CM frame,
\be
g_V^i = T^i_3 - 2 \,Q_i \sin^2 \theta_W\, , \quad g_A^i = T^i_3\, ,
\ee
 and
\be
\Re\big[\chi(q^2)\big] =  \frac{q^2(q^2-m_Z^2)}{\sin^2{\theta_W}\cos^2{\theta_W} \left[(q^2-m_Z^2)^2 + q^4 \Gamma_Z^2/m_Z^2\right]}\quad\, ,
\ee
\be
\left|\chi(q^2)\right|^2 =  \frac{q^4}{\sin^4{\theta_W}\cos^4{\theta_W}\left[(q^2-m_Z^2)^2 + q^4\Gamma_Z^2/m_Z^2\right]}\, ,
\ee
where $\theta_W$ is the Weinberg angle, $m_Z$ and $\Gamma_Z$ the mass and total width of the $Z$ boson respectively, and $q^2=(q_1+q_2)^2$.

\subsubsection{New physics: Contact interactions}
Here, we write the results  for the coefficients $\tilde{A}^{q\bar{q}}$, $\tilde{B}^{qq}_{i}$, and $\tilde{C}^{qq}_{ij}$  for the contact interaction in \eq{CI}---for the benchmark operator $ \displaystyle \frac{4\pi}{\Lambda^2} ( \bar q_L \gamma^\alpha q_L) \,(\bar \tau_R \gamma_\alpha \tau_R )$---in which we retained only the interference with the corresponding SM amplitudes:
\begin{subequations}
\begin{align}
 \tilde{A}^{q\bar{q}} &=   F_\Lambda\, \bigg\{ Q_q Q_\tau \Big[ \beta_\tau^2 \sin^2\Theta + 2 \beta_\tau  \cos \Theta -2  \Big] 
 +  \Rechi\left( g_V^q + g_A^q\right)
 \Big[-2g_V^{\tau}+2\beta_\tau\left( g_V^{\tau} - g_A^{\tau} \right)\cos \Theta 
\nn \\
&     +\beta_\tau^2 g_A^{\tau}\left(1+\cos^2 \Theta\right)+\beta_\tau^2 g_V^{\tau}\sin^2 \Theta\Big]\bigg\}\, ,
   \\ \nn \\
%%%%%%%%%%%%%%%%%%%%%%%%%%%%%%%%%%%%%%%%%%%%%%%
   \tilde{C}^{q\bar{q}}_{nn} &=   F_\Lambda\, \beta_\tau^2 \sin^2 \Theta
   \bigg\{ Q_q Q_\tau + \Rechi \big(g_V^q + g_A^q \big) \big( g_A^{\tau} + g_V^{\tau} \big) \bigg\}\, ,\\
%%%%%%%%%%%%%%%%%%%%%%%%%%%%%%%%%%%%%%%%%%%%%%%
   \tilde{C}^{q\bar{q}}_{rr} &=F_\Lambda\, \sin^2 \Theta
   \bigg\{ Q_q Q_\tau\left[\beta_\tau^2-2\right] + \Rechi \big( g_V^q + g_A^q \big)
   \Big[ \beta_\tau^2\left(g_V^{\tau}-g_A^{\tau}\right)
     -2g_V^{\tau} \Big] \bigg\}\, , \\
%%%%%%%%%%%%%%%%%%%%%%%%%%%%%%%%%%%%%%%%%%%%%%%
   \tilde{C}^{q\bar{q}}_{kk} &=F_\Lambda
   \bigg\{ Q_q Q_\tau \Big[2\beta_\tau\cos \Theta -2\cos^2 \Theta-\beta_\tau^2\sin^2 \Theta\Big]
+ \Rechi \big( g_V^q + g_A^q \big)
\Big[ -2g_V^{\tau}\cos^2 \Theta
\nn \\
& +\beta_\tau\left(g_V^{\tau}-g_A^{\tau}\right)
\left(2\cos \Theta -\beta_\tau\right) +
  \beta_\tau^2 \left(g_A^{\tau}+g_V^{\tau}\right)\cos^2 \Theta\Big] \bigg\}\, , \\
%%%%%%%%%%%%%%%%%%%%%%%%%%%%%%%%%%%%%%%%%%%%%%%%%%%%%%%%%%%%%%%%%%%%%%%%
\tilde{C}^{q\bar{q}}_{kr} &=\tilde{C}^{q\bar{q}}_{rk} = F_\Lambda\, \sqrt{1-\beta_\tau^2}\sin \Theta
   \bigg\{ Q_q Q_\tau\Big[\beta_\tau-2\cos\Theta\Big] \nn \\ & + \Rechi \big( g_V^q + g_A^q \big)
   \Big[  \beta_\tau\big(g_V^{\tau}-g_A^{\tau})-2 g_V^{\tau}\cos \Theta \Big] \bigg\}\, , \\
   %%%%%%%%%%%%%%%%%%%%%%%%%%%%%%%%%%%%%%%%%%%%%%%%%%%%%%%%%%%%%%%%%
        \tilde{C}^{q\bar{q}}_{rn} &=  \tilde{C}^{q\bar{q}}_{nr}=
     \tilde{C}^{q\bar{q}}_{kn}=\tilde{C}^{q\bar{q}}_{nk}=0\, ,
   \\ \nn \\
%%%%%%%%%%%%%%%%%%%%%%%%%%%%%%%%%%%%%%%%%%%%%%%%%%%%%%%%%%%%%%%%%%%%%
   \tilde{B}^{q\bar{q}}_{k} &= F_\Lambda
   \bigg\{ Q_q Q_\tau\Big[2\cos\Theta-\beta_\tau\left(1+\cos^2\Theta\right)\Big] + \Rechi \big( g_V^q + g_A^q \big)
   \Big[-2\beta_\tau^2 g_A^{\tau}\cos\Theta
\nn \\
&     +\beta_\tau\big(g_A^{\tau}-g_V^{\tau})\left(1+\cos^2\Theta\right)
+2g_V^{\tau}\cos \Theta  \Big] \bigg\}\, , \\   
%%%%%%%%%%%%%%%%%%%%%%%%%%%%%%%%%%%%%%%%%%%%%%%%%%%%%%%%%%%%%%%%%
   \tilde{B}^{q\bar{q}}_{r} &=F_\Lambda\, \sqrt{1-\beta_\tau^2} \sin \Theta
   \bigg\{ Q_q Q_\tau\Big[2-\beta_\tau\cos\Theta\Big] + \Rechi \big( g_V^q + g_A^q \big)
   \Big[\beta_\tau\left(g_A^{\tau}-g_V^{\tau}\right)\cos\Theta+2g_V^{\tau} \Big] \bigg\}\, , \\
%%%%%%%%%%%%%%%%%%%%%%%%%%%%%%%%%%%%%%%%%%%%%%%%%%%%%%%%%%%%%%%%%%%%%%%
      \tilde{B}^{q\bar{q}}_{n} &= 0\, ,
\end{align}
where $\displaystyle F_\Lambda = \frac{\mtautau^2}{8 \alpha \Lambda^2}$.
\end{subequations}

%%%%%%%%%%%%%%%%%%%%%%%%%%%%%%%%%%%%%%

%%%%%%%%%%%%%%%%%%%%%%%%%%%%%%%%%%%%%%%%%%%%%%

\end{document}